\documentclass[pra, twocolumn, aps]{revtex4-1}
\usepackage{graphicx}
\usepackage{amsmath,amssymb,dsfont,bbold}
\usepackage{hyperref}

\newcommand{\axis}[1]{\mathbf{\hat #1}}

\begin{document}
\title{Efficient Two-Qubit Pulse Sequences Beyond CNOT}

\author{Daniel Zeuch$^1$ and N.E.~Bonesteel$^2$}

\affiliation{$^1$Peter Gr\" unberg Institut, Theoretical Nanoelectronics, Forschungszentrum  J\" ulich, D-52425 J\"ulich, Germany \\
$^2$Department of Physics and National High Magnetic
Field Laboratory, Florida State University, Tallahassee, FL 32310,
USA}

\date{\today}

\begin{abstract}
We design efficient controlled-rotation gates with arbitrary angle acting on three-spin encoded qubits for exchange-only quantum computation.  Two pulse sequence constructions are given.  The first is motivated by an analytic derivation of the efficient Fong-Wandzura sequence for an exact \textsc{cnot} gate.  This derivation, briefly reviewed here, is based on elevating short sequences of \textsc{swap} pulses to an entangling two-qubit gate.  To go beyond \textsc{cnot}, we apply a similar elevation to a modified short sequence consisting of \textsc{swap}s and one pulse of arbitrary duration.  This results in two-qubit sequences that carry out controlled-rotation gates of arbitrary angle.  The second construction streamlines a class of arbitrary \textsc{cphase} gates established earlier.  Both constructions are based on building two-qubit sequences out of subsequences with special properties that render each step of the construction analytically tractable.
\end{abstract}

\maketitle

\section{Introduction}

In their original proposal for spin-based quantum computation, Loss and DiVincenzo envisioned a quantum computer in which every qubit is encoded into the two-dimensional Hilbert space of a spin-$\frac12$ particle, such as the spin of an electron trapped in a quantum dot \cite{loss98}.  Universal quantum gates can be carried out by adiabatically pulsing the Heisenberg exchange Hamiltonian $J \mathbf S_i \cdot \mathbf S_j$ between pairs of spins, if combined with controlling time-dependent magnetic fields that are local in the spin positions.  In a subsequent proposal, in which logical qubits are represented by pairs of spin-$\frac12$ particles, universality is achieved through control of the Heisenberg exchange Hamiltonian supplemented by a static magnetic field gradient within each two-spin qubit \cite{levy02}.  If, however, logical qubits are encoded using at least three spin-$\frac12$ particles, controlled exchange by itself is a universal resource for quantum computation \cite{bacon00,kempe01}.  

A comprehensive review of computing schemes based on various three-spin qubit implementations is given in Ref.~\cite{russ17}.  In the present work we focus on the exchange-only proposal introduced in Ref.~\cite{divincenzo00}, in which each qubit is encoded using three spin-$\frac12$ particles, and quantum gates are realized by turning on-and-off the exchange coupling between pairs of spins.  As opposed to alternate exchange-only computing schemes in which the exchange coupling acting between certain spins is always turned on \cite{weinstein05, shi12, koh12, taylor13, doherty13, de15, shim16, wardrop16, wang17, ladd17}, in the present computing scheme this coupling is assumed to be completely off unless it is being pulsed.  Reference \cite{divincenzo00} provided the first explicit exchange-pulse sequences forming a universal gate set consisting of arbitrary single-qubit rotations and a controlled-\textsc{not} (\textsc{cnot}) gate (see also Ref.~\cite{kawano05}).  

After the first demonstration of coherent control of the exchange interaction in a semiconductor double quantum dot \cite{petta05}, there have been numerous advances in fabricating and operating quantum dot systems for three-spin qubits \cite{laird10, gaudreau12, medford13_nn, medford13_prl, kim14, eng15, reed16, cao16, thorgrimsson17, andrews19}.  Among these experiments are many realizations of the particular kind of exchange-only qubit considered here \cite{laird10, gaudreau12, medford13_nn, eng15, reed16}, including a recent demonstration of a single-qubit device with average gate errors of 0.35\% \cite{andrews19}.  

For exchange-only quantum computation with exchange pulses, single-qubit rotations are conceptually easy to obtain, requiring at most four pulses for an arbitrary rotation about the Bloch sphere \cite{divincenzo00}.  The construction of two-qubit gates is significantly more complicated.  This is because interqubit pulses, which act on spins that belong to different logical qubits, cause leakage out of the encoded qubit space, while any pulse sequence resulting in a logical gate needs to maintain the qubit encoding.  Given the large search space for unitary operators acting on six spins, the necessity of such interqubit pulses greatly complicates the problem of finding pulse sequences for entangling two-qubit gates.  Indeed, most such sequences have first been found through numerical searches \cite{divincenzo00, hsieh03, fong11, setiawan14}, among which is the optimal known two-qubit sequence found by Fong and Wandzura, which results in a \textsc{cnot} gate \cite{fong11, optimal}.  Recently, we provided a straightforward, analytic derivation of the Fong-Wandzura pulse sequence \cite{zeuch16}.  

The first two-qubit gate sequences that have originally been constructed in an analytic fashion are those presented in Ref.~\cite{zeuch14}.  These sequences can be used to directly enact an arbitrary Controlled-\textsc{phase} (\textsc{cphase}) gate, which, when acting on two qubits with state labels $a=0$ or 1 and $b=0$ or 1, is defined as applying the identity to the two-qubit states $|ab\rangle = |00\rangle$, $|01\rangle$ and $|10\rangle$ while multiplying the $|11\rangle$ state by a phase factor of $e^{i\phi}$ for some phase $\phi$.  Note that such \textsc{cphase} gates with small phases $\phi$ are an integral part in (i) the standard implementation of the quantum Fourier transform (see, e.g., Sec.~5.1 in Ref.~\cite{nielsen10}), and (ii) variational quantum eigensolver algorithms \cite{peruzzo14, mcclean16, yuan19} applied to certain chemistry applications \cite{alexeev19}.  The ability to carry out efficient \textsc{cphase} gates may thus prove valuable in NISQ devices \cite{preskill18}.  As opposed to those pulse sequences mentioned above \cite{divincenzo00, hsieh03, fong11, setiawan14}, the sequences of Ref.~\cite{zeuch14} have a built-in degree of freedom that allows one to directly carry out arbitrary \textsc{cphase} gates.  

Recently, another set of pulse sequences for \textsc{cnot} has been constructed using analytic tools rather than brute force search \cite{van19}.  A crucial idea in this interesting development is to simplify the search for two-qubit sequences by allowing for leakage out of the encoded Hilbert space.  The amount of leakage is then systematically suppressed by iteratively projecting these operations onto the computational Hilbert space, with every subsequent iteration reducing the amount of leakage while increasing the length of the pulse sequence.  Under certain assumptions, some of the sequences presented in Ref.~\cite{van19} have even shorter total duration (though comprising more clock cycles) than the Fong-Wandzura sequence.  

In this paper we show in two different two-qubit gate constructions how the insights gained in our earlier work on deriving two-qubit gate sequences can be used to analytically construct related sequences resulting in controlled-rotation gates locally equivalent to arbitrary \textsc{cphase}.  To make the derivation of our sequences intuitively accessible, we review some of the core aspects of the relevant previous studies \cite{zeuch14, zeuch16}.  The sequences found in this work, similar to those of Ref.~\cite{zeuch14}, contain a small number of exchange pulses whose durations can be adjusted to choose the phase $\phi$ of the \textsc{cphase} gate.  

In the first construction, we design an entirely new class of two-qubit gate sequences by modifying the fundamental structure of the Fong-Wandzura sequence, which, as we have shown previously, manifests itself as a simple three-spin sequence of five exchange pulses \cite{zeuch16}.  We also substantiate this result by working out a concrete member of this gate class.  In the second construction, we take a set of tools developed in an earlier work on designing two-qubit gate sequences \cite{zeuch14} and reorganize them in a way that significantly reduces the total pulse count at the cost of some additional complexity in the construction.  
Assuming the spins encoding the logical qubits are arranged along a linear array and only nearest-neighbor pulses are allowed, the arbitrary \textsc{cphase} sequences presented here consist of only a few more pulses than the optimal Fong-Wandzura \textsc{cnot} sequence.  Our two-qubit gates are efficient, since building arbitrary \textsc{cphase} out of \textsc{cnot} and single-qubit rotations requires calling \textsc{cnot} twice.  As detailed in the Conclusions \ref{conclusions}, compared to this double-\textsc{cnot} construction or the \textsc{cphase} pulse sequences constructed in Ref.~\cite{zeuch14}, we find that our new sequences reduce the total pulse count by a factor $\sim1.5$.  

This paper is organized as follows.  Section \ref{hilbert} discusses some of the basic tools and notation used in this study.  In Sec.~\ref{22}, we present the construction---based on our earlier derivation of the Fong-Wandzura sequence \cite{zeuch16}---of conceptually new two-qubit gates.  Next, in Sec.~\ref{23} we establish two-qubit sequences by way of refining the earlier arbitrary-\textsc{cphase} construction presented in Ref.~\cite{zeuch14}.  Explicit example pulse sequences for arbitrary \textsc{cphase} gates, one for each construction, are given in Sec.~\ref{explicit}, and we conclude in Sec.~\ref{conclusions}.  The equivalence of the two different representations of the Fong-Wandzura sequence given in Refs.~\cite{fong11} and \cite{zeuch16} is worked out explicitly in Appendix \ref{rearrange}.  Appendices \ref{TS} through \ref{explicit_calculations} provide various supporting explanations and calculations.

\section{Rotational Symmetry}
\label{hilbert}

In this work we specify states of multiple spins using only total-spin quantum numbers.  We are allowed to do so because the isotropic Heisenberg exchange Hamiltonian $J \mathbf S_i \cdot \mathbf S_j$, the only resource for realizing quantum gates considered here, is rotationally invariant.

\begin{figure}
	\includegraphics[width=\columnwidth]{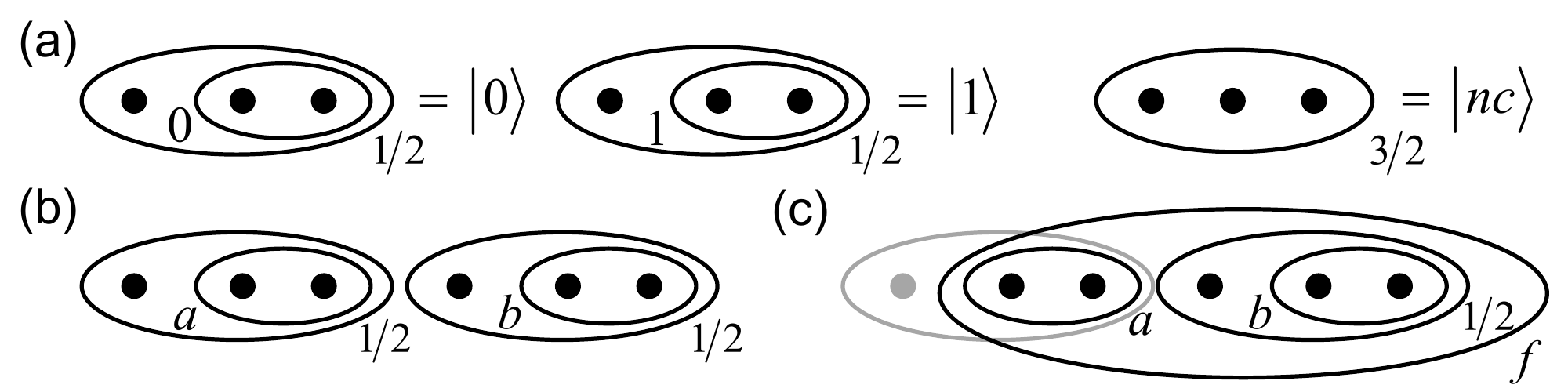}
	\caption{Various multispin states given in a notation in which spin-$\frac12$ particles, represented by $\bullet$, are enclosed by ovals labeled by total spin.  (a) Three-spin qubit encoding of Ref.~\cite{divincenzo00} showing the logical $|0\rangle$ and $|1\rangle$ states, together with the noncomputational state, $|nc\rangle$.  (b) A pair of three-spin qubits with state labels $a$ and $b$.  (c)  Highlighting the five rightmost spins with total spin $f=\frac12$ or $\frac32$ (but not $\frac52$, because the rightmost three spins are initialized with total spin $\frac12$) acted on by the two-qubit pulse sequences constructed in this paper.}
	\label{qubits}
\end{figure}

Figure \ref{qubits} shows the three-spin qubit encoding of Ref.~\cite{divincenzo00} in a convenient notation (introduced to the present context in Ref.~\cite{zeuch14}) in which spins are enclosed by ovals labeled by the total spin of all particles inside.  In this encoding, which we adopt in our work, logical qubits are represented by three spins with total spin $\frac12$.  In the text a spin-$\frac12$ is represented by the symbol $\bullet$, and spins are enclosed by parentheses labeled by total spin.  For instance, the three-spin qubit basis states shown in Fig.~\ref{qubits}(a) are written in the text as $|a\rangle=(\bullet(\bullet\bullet)_a)_{1/2}$ with $a=0$ or 1 defining the standard qubit basis.  Spin states---ordered top to bottom in all following figures---are ordered left to right in the text.  Figure \ref{qubits}(a) also shows the noncomputational three-spin state denoted by $|nc\rangle$, which has total spin $\frac32$.  Note that we treat a spin state like $(\bullet(\bullet\bullet)_a)_{1/2}$ as a single state in Hilbert space even though it is two-fold degenerate taking into account $S_z$ quantum numbers.  

\begin{figure*}
	\includegraphics[width=\textwidth]{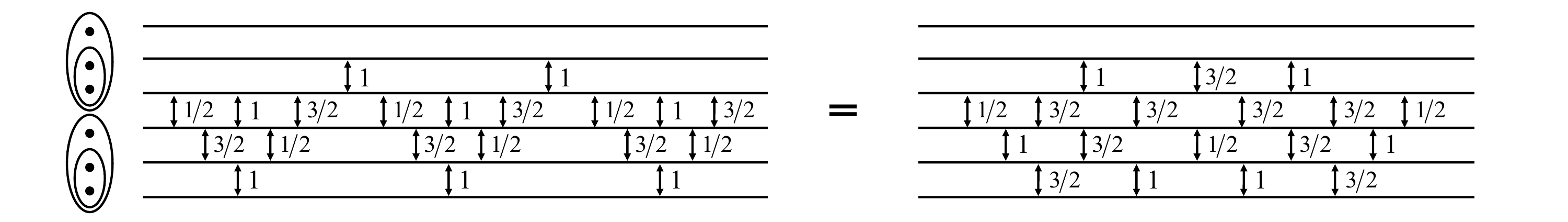}
	\caption{Two representations of the Fong-Wandzura pulse sequence, which are locally equivalent (or equal up to a single-qubit operation).  In Appendix \ref{rearrange} it is shown explicitly how the sequence on the LHS, derived in Ref.~\cite{zeuch16}, can be turned into the sequence on the RHS, obtained numerically in Ref.~\cite{fong11} (given here up to single-qubit rotations).}
	\label{FWsequences}
\end{figure*}

Suppose that we pulse the exchange Hamiltonian $H = J (\mathbf{S}_i \cdot \mathbf{S}_j + \frac34)$ acting on two spins.  In the basis $(\bullet\bullet)_a$ with state ordering $a=\{0,1\}$, the matrix representation of the time evolution operator corresponding to such a pulse, $U_{ij}(t)$ with dimensionless time $t\in[0,2)$ (setting $\hbar=1$), is given by
\begin{eqnarray}
	U_{ij}(t) &=& \text{diag}(1,e^{-i\pi t})
	\label{pulse}\\
			  &=& e^{-i\pi t/2}e^{i\pi t\hat{\bf z}\cdot \boldsymbol\sigma/2}.
	\label{pulsePseudo}
\end{eqnarray}
Here, $\boldsymbol\sigma=(\sigma_x,\sigma_y,\sigma_z)$ denotes the Pauli matrix vector.  Note that we work in units with $\pi/J=1$ so that $t=1$ carries out a \textsc{swap} up to an overall phase.  The inverse of an exchange pulse of duration $t$ is a pulse of duration $s=2-t$.  We often use the fact that, if we take the states $(\bullet\bullet)_{0}$ and $(\bullet\bullet)_{1}$ as the up and down states of a pseudospin, Eq.~(\ref{pulsePseudo}) can be interpreted as a $z$-axis pseudospin rotation through positive angle $\pi t$ [up to an overall phase with respect to the standard SU(2) phase choice].  

Below we construct sequences of exchange pulses, which are described by time evolution operators of the type of Eq.~(\ref{pulse}), for carrying out entangling two-qubit gates.  As pointed out above, finding such a two-qubit sequence is nontrivial because pulse sequences acting on two three-spin qubits do not, in general, maintain the qubit encoding.  To be specific, consider the two encoded qubits shown in Fig.~\ref{qubits}(b).  Note that exchange pulses applied to spins within a three-spin qubit leave its total spin invariant, whereas a pulse acting on two spins from different encoded qubits (in general) alters each qubit's total spin and therefore results in leakage out of the computational space.  

The two-qubit gate sequences presented below can be applied to only five of the six spins encoding two logical qubits, which can be chosen to be those five spins highlighted in Fig.~\ref{qubits}(c).  As shown in Ref.~\cite{zeuch14}, any pulse sequence that acts on only five spins and carries out a leakage-free two-qubit gate for total spin 1 of the two encoded qubits results in the very same gate for total spin 0.  In contrast, when applying the original \textsc{cnot} sequence found by DiVincenzo \emph{et al}. \cite{divincenzo00} to the qubits shown on the LHS of Fig.~\ref{qubits}(b), all six spins undergo nontrivial exchange pulses (i.e., pulses different from \textsc{swap}), and a \textsc{cnot} gate is carried out only if the total spin of the two encoded qubits is 1.

\section{Generalizing the Derivation of the Fong-Wandzura Sequence}
\label{22}

In this section we construct a family of two-qubit gate sequences whose derivation can be understood as a generalization of the derivation of the Fong-Wandzura pulse sequence presented in Ref.~\cite{zeuch16}.  It is thus worthwhile reviewing some of the main steps of that derivation (see also Note \cite{MM2020}).  In doing this, we take the Fong-Wandzura sequence as a starting point and imagine ``reverse engineering'' it to reveal its fundamental structure, and then show how this structure can be altered to find new two-qubit gate sequences.  We note that the order of ideas presented in this review is the reverse of that of Ref.~\cite{zeuch16}, where the Fong-Wandzura sequence is constructed essentially from the bottom up.  

The derivation given in Ref.~\cite{zeuch16} is based on the observation that the pulses of the Fong-Wandzura sequence as published in Ref.~\cite{fong11} can be rearranged without changing the unitary operation carried out by this sequence so that it consists of recurring patterns of a smaller pulse sequence.  This fact is illustrated in Fig.~\ref{FWsequences}, where we show two different representations of the Fong-Wandzura sequence acting on two encoded three-spin qubits.  The LHS of Fig.~\ref{FWsequences} shows a pulse sequence, which was analytically derived in Ref.~\cite{zeuch16} as an equivalent representation of the originally published version of the Fong-Wandzura sequence \cite{fong11}, which, upon removing four pulses used for single-qubit rotations, is shown on the RHS of the same figure.  In Appendix \ref{rearrange} it is shown explicitly how the sequences of Fig.~\ref{FWsequences} can be turned into one another (up to a single-qubit \textsc{swap} pulse not shown in this figure) using only a small set of elementary manipulations, such as moving \textsc{swap} pulses past other pulses, or combining neighboring pulses that act on the same pair of spins to single pulses.  

\begin{figure} 
	\includegraphics[width=\columnwidth]{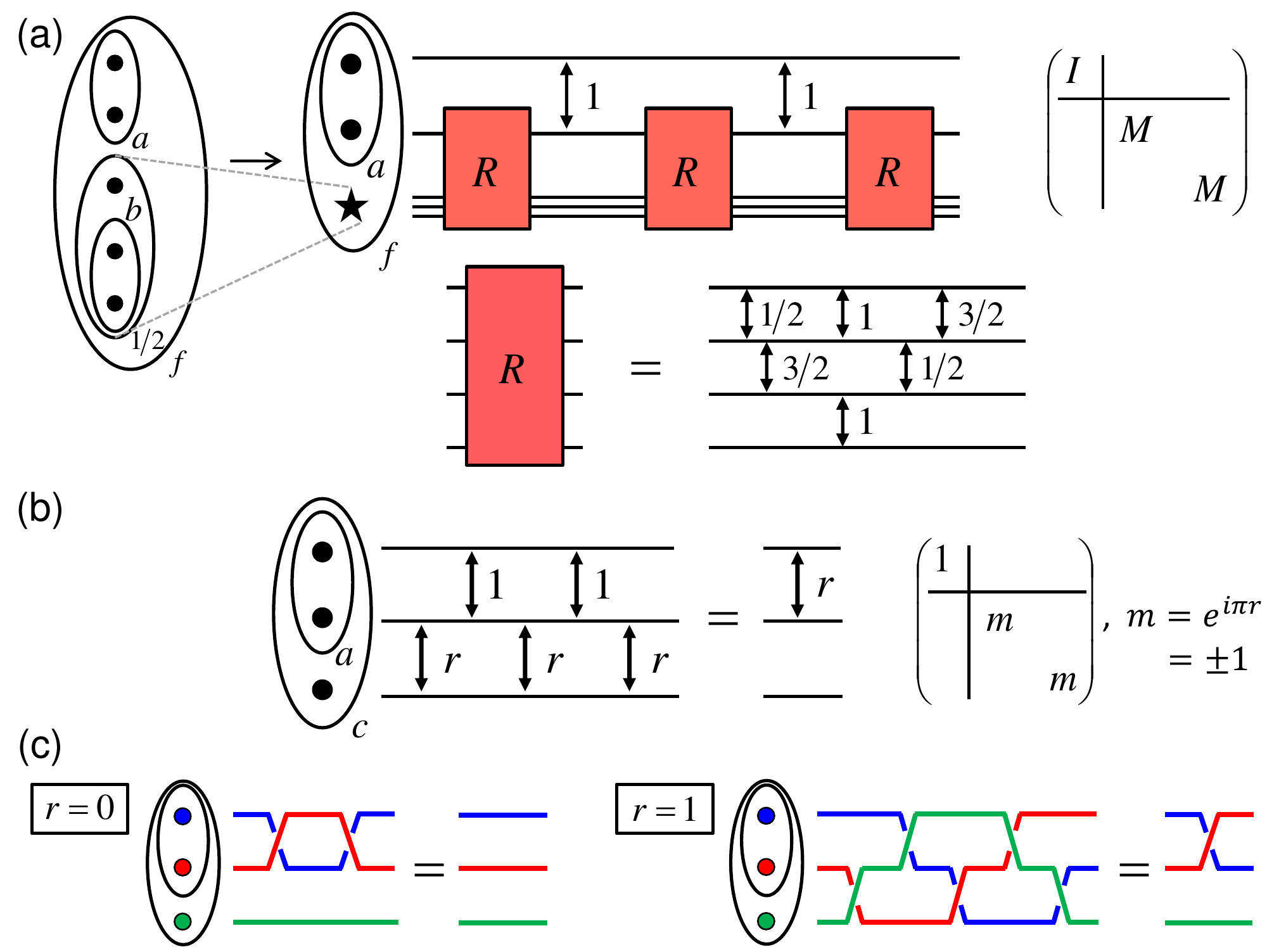}
	\caption{(Color online)  Crucial step in our derivation of the Fong-Wandzura sequence \cite{zeuch16}.  (a) The Fong-Wandzura sequence, as given on the LHS of Fig.~\ref{FWsequences}, applied to the five spins shown in Fig.~\ref{qubits}(c), together with its matrix representation in the $af$-basis (\ref{effectivebasis}) with $M = \hat n_0\cdot \boldsymbol{\sigma}$.  The lowermost three spins are represented by an effective spin-$\frac12$ particle, $\bigstar$, and $R$ represents the repeated six-pulse sequence.  (b) Simple three-spin sequence consisting of three $r$-pulses (pulses with duration $r=0$ or 1 as defined in the main text) and two explicit \textsc{swap}s, which is used to deduce crucial properties of the sequence in (a).  (c) Graphical evaluation of the sequence shown in (b) for the cases $r=0$ and 1, where \textsc{swap} pulses are represented by particle permutations.  As described in the text, we use (c) to show that the sequence in (b) is, in both cases, equivalent to an $r$-pulse applied to the top two spins.  In (b) we also show the matrix representation of this $r$-pulse in the basis $ac=\{0\frac12| 1\frac12, 1\frac32\}$.}
	\label{FWobservation1}
\end{figure}

Figure \ref{FWobservation1} illustrates the main steps of our reverse-engineering process.  In the version of the Fong-Wandzura sequence shown on the LHS of Fig.~\ref{FWsequences}, a six-pulse sequence appears three times.  As shown in Fig.~\ref{FWobservation1}(a), we refer to this repeated sequence as $R$.  It has been shown in Ref.~\cite{zeuch16} that $R$ preserves the total spin of the three spins encoding the logical qubit with state label $b$.  This qubit therefore suffers no leakage when acted on by $R$ or, consequently, by the full \textsc{cnot} sequence.  

As shown in Fig.~\ref{FWobservation1}(a), we represent the three spins encoding the qubit with state label $b$ by what we call an effective spin-$\frac12$ particle $\bigstar$,
\begin{equation}
	(\bullet(\bullet\bullet)_b)_{1/2} =  \bigstar.
	\label{bigstar}
\end{equation}
This allows us to view the Hilbert space of the five spins shown in the figure as a tensor product of the Hilbert space of three spin-$\frac12$ particles with that of a single qubit.  Ignoring $S_z$ quantum numbers, the Hilbert space of these five spins is spanned by the states $((\bullet\bullet)_a(\bullet(\bullet\bullet)_b)_{1/2})_f$ with $ab=00$, 01, 10 and 11 if $f=\frac12$, together with $ab=10$ and 11 if $f=\frac32$.  When replacing the three rightmost spins $(\bullet(\bullet\bullet)_b)_{1/2}$ by $\bigstar$, for the state labeling that corresponds to the basis of the matrix representation of the Fong-Wandzura sequence in Fig.~\ref{FWobservation1}(a) we have
\begin{eqnarray}
	abf = \{00\frac12,01\frac12|10\frac12,11\frac12,10\frac32,11\frac32\} \qquad\qquad\qquad \nonumber\\
		\longrightarrow \quad af=\{\mathbf{0\frac12}|\mathbf{1\frac12},\mathbf{1\frac32}\}. \quad
\label{effectivebasis}
\end{eqnarray}
Note that this $af$ basis is spanned by two-dimensional basis states, printed in bold face, and that the matrix shown in Fig.~\ref{FWobservation1}(a) consists of $2\times2$ block elements rather than numbers, reflecting the tensor product structure described above.  Unless otherwise noted, we adopt the convention that, as in the basis (\ref{effectivebasis}) or the matrices shown in Fig.~\ref{FWobservation1}, solid lines separate Hilbert space sectors with $a=0$ and 1.  The reason we are allowed to concentrate on the Hilbert space spanned by the basis (\ref{effectivebasis}) is that, as emphasized above, $R$ conserves the total spin of the logical qubit represented by $\bigstar$.  To continue using this $af$ basis, in the present section we require all pulse sequences that act on a collection of spins including $\bigstar$ to similarly conserve the total spin of $\bigstar$.  

Given the matrix shown in Fig.~\ref{FWobservation1}(a), the operation carried out by the two-qubit sequence shown in this figure applies the $2\times2$ identity, $\mathbb{1}$, if $a=0$, and the $2\times 2$ matrix $M=\hat{\bf n}_0\cdot\boldsymbol\sigma$ \cite{zeuch16} if $a=1$, regardless of the value of $f$; here $\hat{\bf n}_0$ is a certain three-dimensional unit vector (see Note \cite{n0}).  Accordingly, in the two-qubit basis $ab=\{00,01,10,11\}$ the matrix of the gate carried out by this sequence is
\begin{equation}
	U_{\text{FW}}=\text{diag}(\mathbb{1},M).
	\label{FW}
\end{equation}
With $M=\hat{\bf n}_0\cdot\boldsymbol\sigma$ this gate is locally equivalent to \textsc{cnot}.  

In Ref.~\cite{zeuch16}, the two-qubit sequence of Fig.~\ref{FWobservation1}(a) has been derived through an elevation of the simpler sequences shown in Fig.~\ref{FWobservation1}(b).  By elevation we mean a process of deriving five-spin sequences from simpler three-spin sequences in a way that allows us to infer properties of the unitary operator produced by the former based on the latter.  In this case the simpler sequences consist of explicit \textsc{swap}s and pulses of duration $r=0$ or 1, which we denote as $r$-pulses.  According to Eq.~(\ref{pulse}), the matrix representation of an $r$-pulse acting on two spins in the basis $(\bullet\bullet)_a$ with state ordering $a=\{0,1\}$ is 
\begin{eqnarray}
	U_r = \text{diag}(1,e^{-i\pi r})=
	\text{diag}(1,m), \quad m^2 = 1.
	\label{r}
\end{eqnarray}
If $r=0$ then $m=1$ and the $r$-pulse carries out the identity operation.  If $r=1$ then $m=-1$ and the $r$-pulse carries out a \textsc{swap}.  Note that the sequence in Fig.~\ref{FWobservation1}(a) can be obtained by replacing the lowermost spin in Fig.~\ref{FWobservation1}(b) with the particle $\bigstar$ introduced above in Eq.~(\ref{bigstar}), and further replacing each $r$-pulse with an $R$ sequence.  

The pulse sequence shown in Fig.~\ref{FWobservation1}(b) can be evaluated straightforwardly, because all its pulses are equivalent to simple particle permutations (the identity for $r=0$ and \textsc{swap} for $r=1$).  This evaluation is illustrated in Fig.~\ref{FWobservation1}(c) for both $r=0$ and $r=1$, thus verifying the identity that the five-pulse sequence shown in Fig.~\ref{FWobservation1}(b) is equal to a single $r$-pulse applied to the top two spins.  The corresponding matrix representation, shown in Fig.~\ref{FWobservation1}(b), can be directly read off Eq.~(\ref{r}).  

The matrix representation of an initially generic $R$ sequence in the basis $(\bullet\bigstar)_d$ with state ordering $d=\{\mathbf{0},\mathbf{1}\}$ is \cite{zeuch16}
\begin{eqnarray}
	R = \text{diag}(\mathbb{1},M), \quad M^2 = \mathbb{1}.
	\label{R}
\end{eqnarray}
Here, the requirement $M^2 = \mathbb{1}$ is an elevated version of the requirement $m^2=1$ in Eq.~(\ref{r}), and is needed for the $R$ sequence to be viewed as an elevated $r$-pulse.  For the one-dimensional case the equation $m^2=1$ has the two solutions $m=\pm1$.  In contrast, the equation $M^2=\mathbb{1}$ has, in addition to the equivalent solutions $M=\pm\mathbb{1}$, a continuum of solutions, $M=\hat {\bf n}\cdot \boldsymbol\sigma$, where $\hat {\bf n}$ can be any unit vector.  Finally, as discussed in Ref.~\cite{zeuch16}, by elevating the simple pulse sequence of Fig.~\ref{FWobservation1}(b) to that of Fig.~\ref{FWobservation1}(a), one can directly infer the matrix representation of the Fong-Wandzura sequence shown in Fig.~\ref{FWobservation1}(a). 

\begin{figure}
	\includegraphics[width=\columnwidth]{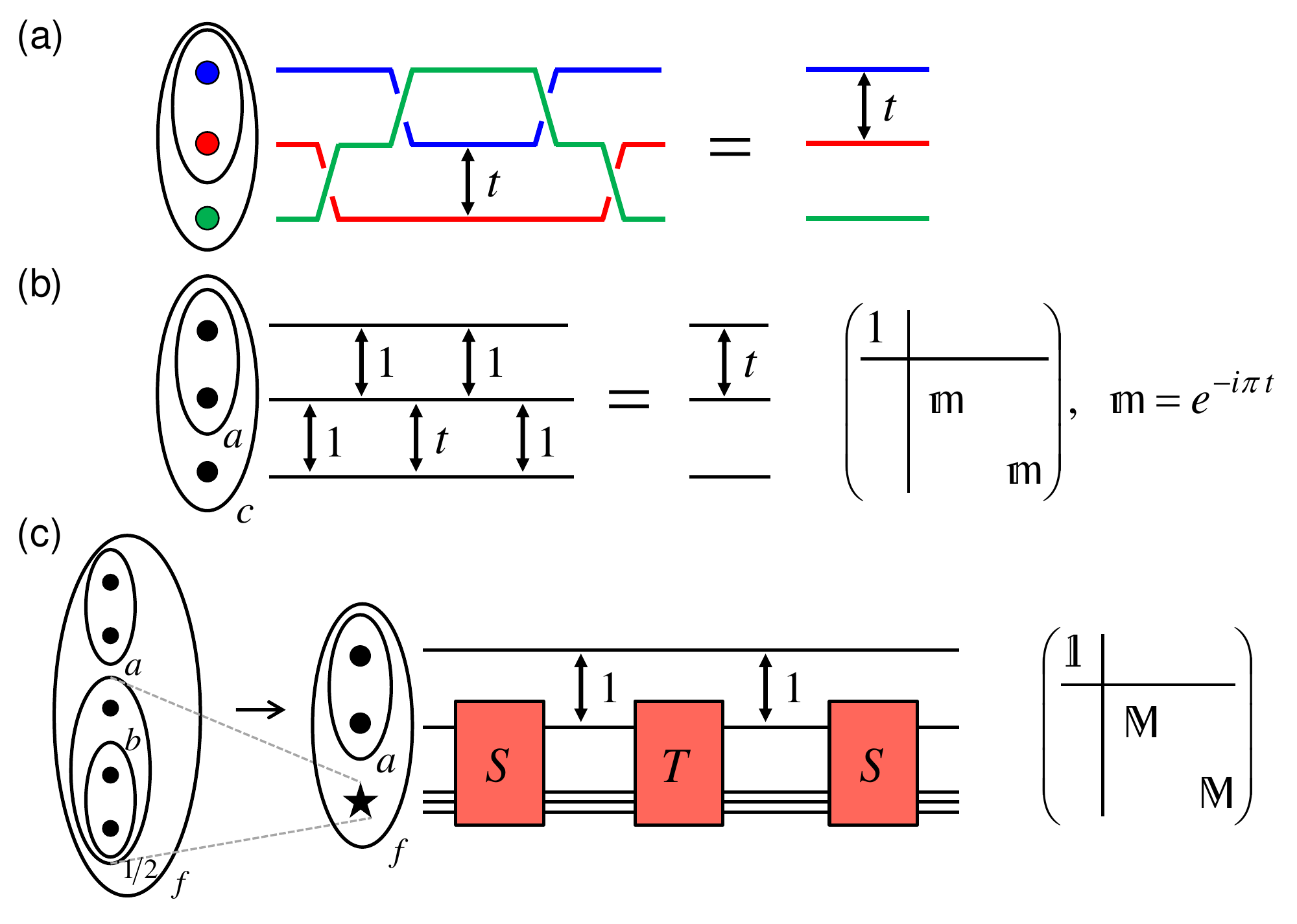}
	\caption{(Color Online)  Two-qubit gate construction generalizing that shown in Fig.~\ref{FWobservation1}, with steps in reverse order.  (a) Sequence of particle interchanges and a $t$-pulse.  It is readily seen that this sequence is equivalent to only applying the $t$-pulse to the top two spins.  (b) Sequence of four \textsc{swap}s and one $t$-pulse, which is simplified using (a) by interpreting \textsc{swap}s as particle interchanges.  Its corresponding matrix representation is shown in the indicated basis with $ac=\{0\frac12|1\frac12,1\frac32\}$.  (c) Two-qubit sequence, which is an elevated version of the sequence shown in (b), applied to the five spins shown in Fig.~\ref{qubits}(c).  The logical qubit (with state label $b$) is represented by an effective spin-$\frac12$, $\bigstar$.  The corresponding matrix with arbitrary $\mathbb M$ is given in the effective basis (\ref{effectivebasis}), i.e., $af=\{\mathbf{0\frac12}|\mathbf{1\frac12},\mathbf{1\frac32}\}$.}
	\label{observation1}
\end{figure}

We now turn to our new pulse sequence construction, which is based on the insights gained through the reverse engineering process described above.  Figure \ref{observation1}(a) shows a simple sequence of spin permutations and one exchange pulse of arbitrary duration $t$, or $t$-pulse, which is clearly equivalent to the single $t$-pulse shown on the RHS.  Replacing all particle permutations with \textsc{swap}s yields the pulse sequence identity shown in Fig.~\ref{observation1}(b).  The matrix representation corresponding to the five-pulse sequence shown in Fig.~\ref{observation1}(b) can thus be directly read off Eq.~(\ref{pulse}).  

In a similar way as the two-qubit pulse sequence shown in Fig.~\ref{FWobservation1}(a) can be interpreted as a generalization of the sequence in Fig.~\ref{FWobservation1}(b), we now generalize the sequence shown in Fig.~\ref{observation1}(b) to the two-qubit sequence in Fig.~\ref{observation1}(c).  This last sequence is applied to the five spins shown in Fig.~\ref{qubits}(c) upon representing the logical qubit with state label $b$ by an effective spin-$\frac12$, $\bigstar$.  The schematic operations $T$ and $S$ can be viewed as elevated versions of $t$-pulse and a \textsc{swap}, respectively, and are realized by pulse sequences---to be determined---which are applied to one spin-$\frac12$, $\bullet$, and the three spins inside the effective spin-$\frac12$, $\bigstar$.  Note that the effective Hilbert space of these particles is spanned by the states $(\bullet\bigstar)_d$ with $d=0$ or 1, but not 2 because the total spin of the three-spin qubit hidden inside $\bigstar$, as given in Eq.~(\ref{bigstar}), is initialized to be $\frac12$.  

The $t$-pulse, whose matrix representation is shown in Fig.~\ref{observation1}(b), is generalized to the $T$ operation by promoting the numbers 1 and $\mathbb{m}=e^{-i\pi t}$ to the unitary $2\times2$ matrices $\mathbb{1}$ and $\mathbb{M}$, respectively.  As opposed to the matrix $M$ introduced above, the matrix $\mathbb M$ is not required to fulfill any special condition (besides being unitary).  The matrix representation of the operation carried out by $T$ when applied as in Fig.~\ref{observation1}(c) in the basis $(\bullet\bigstar)_{d}$ with $d=\{{\bf 0},{\bf 1}\}$ is then
\begin{equation}
	T = \text{diag}(\mathbb{1},\mathbb{M}).
	\label{T}
\end{equation}
A \textsc{swap} operation, whose matrix is given by Eq.~(\ref{r}) for the case of $m=-1$, is similarly generalized to the $S$ operation by promoting the numbers $\pm1$ to the $2\times2$ matrices $\pm\mathbb{1}$, respectively.  Accordingly, the matrix corresponding to the $S$-operation when applied to $\bullet$ and $\bigstar$ as in Fig.~\ref{observation1}(c) in the basis $(\bullet\bigstar)_{d}$ with $d=\{{\bf 0},{\bf 1}\}$ is
\begin{equation}
	S = \text{diag}(\mathbb{1},-\mathbb{1}).
	\label{S}
\end{equation}

The matrix representation of the two-qubit sequence shown in Fig.~\ref{observation1}(c) consists of $2\times2$ block elements that act on the Hilbert space of the three-spin qubit hidden inside the effective spin-$\frac12$, $\bigstar$, because the operations $T$, $S$ and \textsc{swap} conserve this qubit's total spin.  To find this matrix representation, we first note that each of its $2\times2$ block elements must be a polynomial linear in $\mathbb{M}$, $\alpha_0 \mathbb 1 + \alpha_1 \mathbb{M}$, because there is only one operation in this sequence, $T$, whose matrix contains an element unequal to $\pm \mathbb 1$, namely $\mathbb{M}$.  These polynomials are determined for any $\mathbb{M}$ by the special case of $\mathbb{M}=e^{-i\pi t} \mathbb{1}$, for which the sequence shown in Fig.~\ref{observation1}(c) is equivalent to that shown in Fig.~\ref{observation1}(b).  Accordingly, the matrix shown in Fig.~\ref{observation1}(c) is an elevated version of that shown in Fig.~\ref{observation1}(b).

When applying the pulse sequence of Fig.~\ref{observation1}(c) to the logical qubits shown in Fig.~\ref{qubits}(b), we can deduce its action by noting that its matrix representation in the basis (\ref{effectivebasis}) is the identity for $a=0$, and $\mathbb{M}$ for $a=1$ (independent of $f$).  This corresponds to the quantum gate
\begin{equation}
	U_{\text{2qubit}} = \text{diag}(\mathbb{1}, \mathbb{M}).
	\label{2qubit}
\end{equation}
The sequence of Fig.~\ref{observation1}(c) may thus be used to carry out an arbitrary controlled-operation gate with the control and target being the encoded qubits with state labels $a$ and $b$, respectively. For the parameterization
\begin{eqnarray}
	\mathbb{M}(\phi)=e^{i\xi}e^{i \phi \hat{\bf n}\cdot\boldsymbol\sigma/2}
	\label{parameterizeM}
\end{eqnarray}
the two-qubit gate (\ref{2qubit}) is then a controlled-rotation gate with angle $\phi$ and axis $\hat {\bf n}$ and an additional phase $\xi$.  Here we write $\mathbb{M} = \mathbb{M}(\phi)$, since $\phi$ is the only parameter invariant under single-qubit rotations.  

In principle, one can use any four-spin pulse sequences satisfying Eqs.~(\ref{T}) and (\ref{S}) for the operations $T$ and $S$, which have been left implicit up to now.  Since Eq.~(\ref{2qubit}) is a function of $\mathbb M$ only, the actual two-qubit gate then only depends on the particular realization of the $T$ sequence.  In Appendix \ref{TS} we derive an explicit set of pulse sequences for $T$ and $S$.  In doing this, we combine insights gained by deriving the Fong-Wandzura sequence \cite{zeuch16} with tools from Ref.~\cite{zeuch14}.  The resulting set of two-qubit gate sequences is given in Sec.~\ref{explicit}.

\section{Optimized CPhase Gates}
\label{23}

We now present an analytic two-qubit gate construction based on a set of tools and concepts developed previously for constructing pulse sequences for arbitrary \textsc{cphase} gates \cite{zeuch14}.  The resulting sequences carry out the same two-qubit gates using fewer exchange pulses.  

\begin{figure}
	\includegraphics[width=\columnwidth]{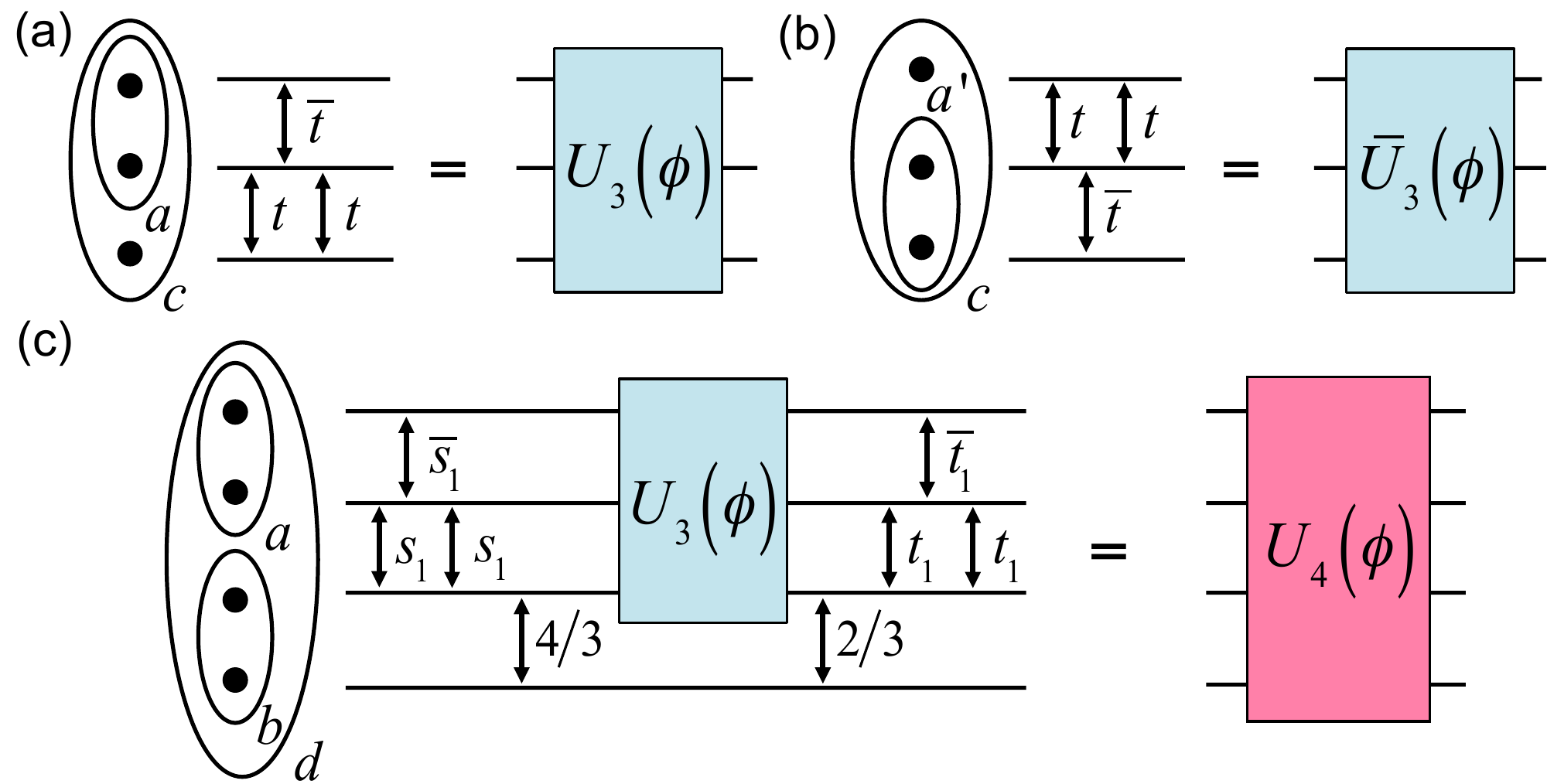}
	\caption{Pulse sequences adapted from Ref.~\cite{zeuch14} carrying out operations serving as building blocks for our two-qubit gate construction.  (a)  Operation $U_3$, whose matrix representation is given by Eq.~(\ref{u3}) in the indicated $ac$-basis.  (b)  Operation $\overline U_3$, whose matrix representation (\ref{u3bar}) is given in the indicated $a'c$-basis $(\bullet(\bullet\bullet)_{a^\prime})_c$.  The parameters $t$, $\bar t$ and $\phi$ shown in both (a) and (b) are related to one another via Eqs.~(\ref{ttbar}) and (\ref{phi}).  (c)  Operation $U_4$, whose matrix representation (\ref{u4}) is given in the indicated $bd$-basis.  Here, the explicit $t_1$, $\bar t_1$, $t_1$ and $s_1$, $\bar s_1$, $s_1$ sequences carry out $U_3(2\pi/3)$ and $U_3(4\pi/3)$ operations, respectively.  By solving Eqs.~(\ref{ttbar}) and (\ref{phi}) for $\phi=2\pi/3$, one finds $t_1=0.426548$ \cite{zeuch14}.  The values of $\bar t_1$, $s_1$ and $\bar s_1$ are then determined by Eq.~(\ref{ttbar}) and the relation $t_1 + s_1 = \bar t_1 + \bar s_1 = 2$, which follows from the fact that $4\pi/3 = 2\pi - 2\pi/3$ (see main text).}
	\label{34}
\end{figure}

Figure \ref{34} shows the three basic pulse sequences used in this construction, $U_3$, $\overline U_3$, and $U_4$.  Consider the three spin-$\frac12$ particles shown in Fig.~\ref{34}(a), whose three-dimensional Hilbert space is spanned by the states $((\bullet\bullet)_a\bullet)_c$ with quantum numbers $ac=0\frac12$, $1\frac12$ and $1\frac32$.  The first building block for our two-qubit gate construction is the operation $U_3$ shown in Fig.~\ref{34}(a), which has been introduced in Ref.~\cite{zeuch14}.  The pulses making up $U_3$ are of durations $t$ and $\bar t$ with $0\leq t \leq \bar t \leq 2$, which fulfill
\begin{eqnarray}
	\tan(\pi t/2)\tan(\pi \bar t/2) = -2.
	\label{ttbar}
\end{eqnarray}
As discussed in Ref.~\cite{zeuch14}, the matrix representation of the operation $U_3$ in the $ac$-basis shown in Fig.~\ref{34}(a) with state ordering $ac=\{0\frac12| 1\frac12, 1\frac32\}$ is
\begin{equation}
	U_3(\phi) = \left(
	\begin{array}{cccc}
		e^{-i \pi \bar t}& \vline\\\hline
		&\vline& 1 					\\
		&\vline&	&	e^{-i \phi}
	\end{array}
	\right),
	\label{u3}
\end{equation}
where
\begin{equation}
	\phi= \pi (t + \bar t - 1), \qquad \ \phi \in [0, 2\pi].
	\label{phi}
\end{equation}
As indicated by the solid lines in Eq.~(\ref{u3}), the operation $U_3$ conserves the total spin of the top two particles (denoted $a$) shown in Fig.~\ref{34}(a).  Note that $U_3$ acts trivially on the one-dimensional $a=0$ subspace, i.e., here it is proportional to the identity, while it applies a phase shift of $e^{-i\phi}$ between the $a=1$ states with $c=\frac12$ and $\frac32$.  

The pulse sequence for $\overline U_3$ is shown in Fig.~\ref{34}(b).  Since this sequence is the mirror image of the $U_3$ sequence in Fig.~\ref{34}(a),  we can directly infer the matrix representation of $\overline U_3$ in the mirrored basis $(\bullet(\bullet\bullet)_{a'})_c$ to be simply that given in Eq.~(\ref{u3}).  Therefore, in the basis $a'c = \{0\frac12| 1\frac12, 1\frac32\}$ we find
\begin{equation}
	\overline U_3(\phi) = \left(
	\begin{array}{cccc}
		e^{-i \pi \bar t}& \vline\\\hline
		&\vline& 1 					\\
		&\vline&	&	e^{-i \phi}
	\end{array}
	\right)
	\label{u3bar}
\end{equation}
with $t$, $\bar t$ and $\phi$ related to one another via Eqs.~(\ref{ttbar}) and (\ref{phi}).  As opposed to the operation $U_3$, this mirrored operation $\overline U_3$ conserves the total spin of the lower two spins shown in Fig.~\ref{34}(b) (denoted $a'$), and applies a phase shift of $e^{-i\phi}$ between the $a'=1$ states while it acts trivially if $a'=0$.  

The inverse of a $t$, $\bar t$, $t$ sequence shown in Fig.~\ref{34}(a), which results in the operation $U_3(\phi)$, is the same three-pulse sequence but with durations $s$, $\bar s$, $s$, where $s = 2-t$ and $\bar s = 2-\bar t$ (note that $s \geq \bar s$ since $t\leq \bar t$).  For the operation carried out by this inverse sequence, $U_3(\chi)$, we take from Eq.~(\ref{phi}) that $\chi = \pi(s + \bar s - 1) = 2\pi - \phi$.  Similarly, the inverse of the $\overline U_3(\phi)$ operation, whose $t$, $\bar t$, $t$ sequence is shown in Fig.~\ref{34}(b), is the same sequence with durations $s$, $\bar s$, $s$ (again with $s = 2-t$ and $\bar s = 2-\bar t$), which carries out the operation $\overline U_3(\chi)$ with the same $\chi=2\pi-\phi$.  %The matrix representations of $U_3(\chi)$ and $\overline U_3(\chi)$ are, respectively, given by Eqs.~(\ref{u3}) and (\ref{u3bar}) in the indicated bases, upon replacing $\phi\rightarrow\chi$ and $\bar t\rightarrow\bar s$.  

Finally, Fig.~\ref{34}(c) shows the third sequence, $U_4$, which is applied to four spins.  By our usual convention of ignoring $S_z$ quantum numbers, the Hilbert space associated with these spins is six-dimensional.  For the indicated basis, $((\bullet\bullet)_a(\bullet\bullet)_b)_d$, the $d=0$ space is spanned by the states with $ab=00$ and 11, the $d=1$ space by the states with $ab=01$, 10 and 11, and the $d=2$ space by the $ab=11$ state.  As shown in the figure, the pulse sequence of $U_4$ consists of a number of pulses of fixed durations, independent of $\phi$, together with a $U_3(\phi)$ sequence [see Fig.~\ref{34}(a)].  

The operation $U_4$ was designed in Ref.~\cite{zeuch14} to apply a simple phase factor to all $a=0$ states $((\bullet\bullet)_{a=0}(\bullet\bullet)_b)_{d=b}$,
\begin{eqnarray}
	\quad U_4(\phi) = e^{-i \pi \bar t}\mathbb{1}, \qquad\qquad (a=0).
	\label{u4a=0}
\end{eqnarray}
We therefore concentrate on the four-dimensional $a=1$ Hilbert space, which is spanned by the states $((\bullet\bullet)_{a=1}(\bullet\bullet)_b)_d$ with $bd = 10$, 01, 11 and 12.  In the basis $bd=\{10|01|11,12\}$ the operation $U_4$ has the matrix representation
\begin{equation}
	U_4(\phi) = \left(
	\begin{array}{cccccc}
		1	& \vline 	& & \vline	  \\ \hline
		  & \vline	& 1&\vline						\\ \hline
		  & \vline	& & \vline & e^{-i \phi}	\\
		  & \vline	& &	\vline & &	e^{-i \phi}
	\end{array}
	\right).
	\label{u4}
\end{equation}
In this matrix, solid lines separate out the one-dimensional $b=0$ sector where $U_4$ acts as the identity.  $U_4$ thus acts trivially for both of the cases $a=0$ or $b=0$.  

Note all three sequences shown in Fig.~\ref{34} act in a simple way.  They (i) conserve the total-spin quantum numbers of certain spin pairs, and (ii) act trivially [i.e., proportional to the identity] only if those quantum numbers are equal to 0.

To design two-qubit gate sequences, we consider the five spins shown in Fig.~\ref{qubits}(c),
\begin{eqnarray}
	((\bullet\bullet)_a(\bullet(\bullet\bullet)_b)_c)_f,
	\label{5spinsBasic}
\end{eqnarray}
where $a, b = 0$ or 1 define the states of the two logical qubits, and $c, f = \frac12$ or $\frac32$.  As opposed to Fig.~\ref{qubits}(c), here we take into account quantum states in which the logical qubit with state label $b$ has total spin $c=\frac32$, because, as explained in Sec.~\ref{hilbert}, the total spin of a three-spin qubit, which is initialized to be $\frac12$, is altered by certain interqubit exchange pulses.  

\begin{figure}
	\includegraphics[width=\columnwidth]{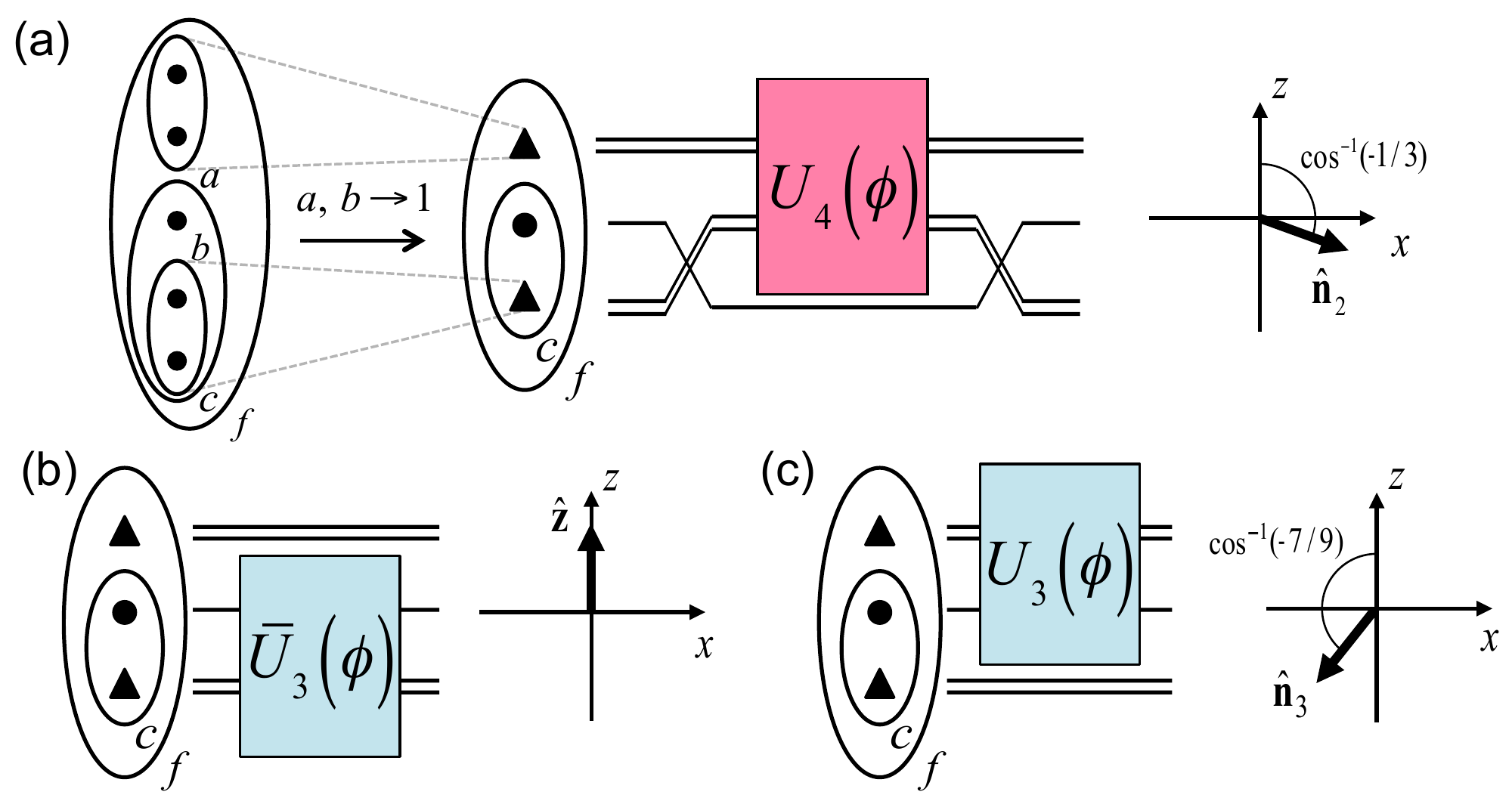}
	\caption{Operations $U_4$, $\overline U_3$ and $U_3$ applied to the five spins shown in Fig.~\ref{qubits}(c).  Each operation acts trivially if the total spin of the topmost or lowermost two particles is 0.  We thus work in an effective Hilbert space by setting both $a$ and $b$ to 1, as indicated by the symbols $\blacktriangle$ [see Eq.~(\ref{introduceBlackTriangle})].  The $U_4$ operation in (a) is surrounded by a permutation of one spin-$\frac12$ particle with two spin-$\frac12$ particles, denoted \textsc{powt}, which ensure that the $U_4$ sequence is applied to the two spin pairs with total spins $a$ and $b$.  Each operation carries out a pseudospin rotation on $\{\uparrow_{1/2}, \downarrow_{1/2}\}$ defined in Eq.~(\ref{pseudospin}) about the indicated axes $\hat {\bf n}_2$ [for $U_4$ as shown in (a)], $\hat {\bf z}$ [for $\overline U_3$ in (b)] and $\hat {\bf n}_3$ [for $U_3$ in (c)].}
	\label{u34}
\end{figure}

Figure \ref{u34} shows how we apply the operations of Fig.~\ref{34} to the spins (\ref{5spinsBasic}).  This particular layout takes full advantage of the spin-conservation feature summarized above, since each of the three operations conserves the quantum numbers $a$ and $b$.  From the discussion above, this conservation is easy to see for the operations $U_3$ and $\overline U_3$.  The $U_4$ operation shown in Fig.~\ref{u34}(a), however, is defined in Fig.~\ref{34}(c) as acting on two neighboring spin pairs with total spins $a$ and $b$.  To ensure the conservation of $b$, the $U_4$ operation is surrounded by permutations of one spin-$\frac12$ particle $\bullet$ with the pair of spins-$\frac12$ particles with total spin $b$ (assumed to be 1 in Fig.~\ref{u34}) represented by $\blacktriangle$.  We refer to this permutation of one spin-$\frac12$ particle with two spin-$\frac12$ particles as \textsc{powt}.  

\begin{figure*}
	\includegraphics[width=\textwidth]{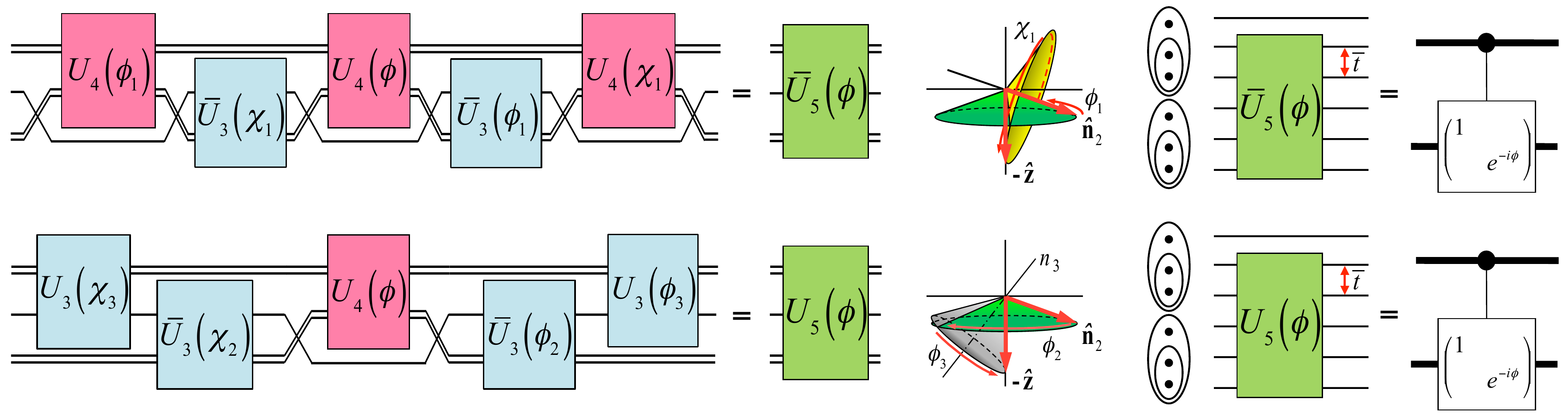}
	\caption{(Color online)  Constructing two-qubit gates applied to the five spins shown in Fig.~\ref{qubits}(c).  Top panel:  $\overline U_5$ gate sequence of Ref.~\cite{zeuch14} consisting of three $U_4$ and two $U_3$-like operations.  Bottom panel:   Optimized sequence $U_5$ consisting of only one $U_4$ and four $U_3$-like operations.  As also shown, $\overline U_5$ and $U_5$, constructed via similarity transformations and augmented by single-qubit pulses explained in the text, can be used to carry out arbitrary \textsc{cphase} gates with $\phi \in [0, 2\pi]$.  The transformation of each sequence is visualized by the respective diagram featuring two intersecting cones where $\chi_i = 2\pi - \phi_i$, and $\phi_i$ for $i = 1, 2, 3$ are given by Eqs.~(\ref{phi1})-(\ref{phi3}).  Note that the cones for the top panel have undergone a $\pi$ rotation about the $z$ axis when compared to those given in Ref.~\cite{zeuch14}; this is due to the change in qubit basis which manifests itself in the six \textsc{powt} operations in this sequence.}
	\label{CPhaseOldNew}
\end{figure*}

We now concatenate the operations shown in Fig.~\ref{u34} to design arbitrary \textsc{cphase} gates of the form
\begin{eqnarray}
	U_{\textsc{cphase}}(\phi) = \text{diag}(1, 1, 1, e^{- i\phi})
	\label{cphase}
\end{eqnarray}
in the two-qubit basis $ab=\{00, 01, 10, 11\}$.  As explained in Appendix \ref{simplification}, the operations of Fig.~\ref{u34} result in simple phase factors if $a=0$ or $b=0$, which only depend on the state of one of the qubits and can therefore be undone by single-qubit rotations before or after the main two-qubit gate sequence.  We can therefore focus on the nontrivial, effective Hilbert space spanned by states (\ref{5spinsBasic}) with $ab=11$.  As also shown in Fig.~\ref{u34}(a), this effective space is obtained by replacing those spin pairs with total spin 1 by effective spin-1 particles denoted $\blacktriangle$,
\begin{eqnarray}
	(\bullet \bullet)_1 &\rightarrow& \blacktriangle,
	\label{introduceBlackTriangle}\\
	((\bullet\bullet)_{a=1}(\bullet(\bullet\bullet)_{b=1})_c)_f & \rightarrow & (\blacktriangle(\bullet\blacktriangle)_c)_f.
	\label{5spins}
\end{eqnarray}
The corresponding four-dimensional Hilbert space is spanned by the states with $cf=\frac12\frac12$, $\frac32\frac12$, $\frac12\frac32$ and $\frac32\frac32$. 

Figure \ref{CPhaseOldNew} shows both the original pulse sequence presented in Ref.~\cite{zeuch14} (top panel, operation $\overline U_5$) as well as the new sequence developed here (bottom panel, operation $U_5$).  [We note that the $\overline U_5$ sequence is a slightly altered version of that published in Ref.~\cite{zeuch14} due to a different choice of qubit bases.]  As shown on the right side of the figure, applying the sequences $\overline U_5$ and $U_5$ to two encoded three-spin qubits enacts an arbitrary \textsc{cphase} gate up to the indicated single-qubit rotations.  Note that, qualitatively speaking, the new sequence for $U_5$ is the same as that for $\overline U_5$ upon replacing the outer $U_4$ operations by $U_3$ operations.  This is why these new sequences contain fewer exchange pulses than those originally published in Ref.~\cite{zeuch14}.  

We now explain how the two \textsc{cphase} gate sequences shown in Fig.~\ref{CPhaseOldNew} can be derived analytically based on geometric intuition.  Our derivation is based on the fact that both $\overline U_5(\phi) = U_4(\chi_1) \overline U_3(\phi_1) U_4(\phi) \overline U_3(\chi_1) U_4(\phi_1)$ and $U_5(\phi) = U_3(\phi_3) \overline U_3(\phi_2) U_4(\phi) \overline U_3(\chi_2) U_3(\chi_3)$ with $\chi_i = 2 \pi - \phi_i$ for $i = 1, 2, 3$ can be understood in terms of similarity transformations.  [Here and below the operations $U_3$, $\overline U_3$, and $U_4$ are understood to be acting on the full five-spin Hilbert space as shown Fig.~\ref{u34}.] To see the similarity transformation structure, note that $\chi_1 = 2\pi-\phi_1$ implies $\overline U_3(\chi_1) = \overline U_3(\phi_1)^{-1}$ and $U_4(\chi_1) = U_4(\phi_1)^{-1}$, so that we can write
\begin{eqnarray} 
	\overline U_5(\phi) &=& P U_4(\phi) P^{-1}, \qquad P = U_4(\chi_1) \overline U_3(\phi_1).
	\label{u5bar}
\end{eqnarray} 
Similarly, $\chi_{2, 3} = 2\pi-\phi_{2, 3}$ implies $\overline U_3(\chi_2) = \overline U_3(\phi_2)^{-1}$ and $U_3(\chi_3) = U_3(\phi_3)^{-1}$, so that
\begin{eqnarray}
	U_5(\phi) &=& Q U_4(\phi) Q^{-1}, \qquad Q = U_3(\phi_3) \overline U_3(\phi_2).
	\label{u5}
\end{eqnarray}

Since each operation in Eqs.~(\ref{u5bar}) and (\ref{u5}) is of the form of a similarity transformation, any phases applied by $P$ or $Q$ on the $a=0$ or $b=0$ subspace cancel out, since here $U_4$ acts proportional to the identity.  For this subspace, the only nonzero phase applied by either $U_5$ or $\overline U_5$ is then that due to $U_4(\phi)$ for $a=0$.  According to Eq.~(\ref{u4a=0}) this phase factor is $e^{-i\pi \bar t}$, which, as shown to the far right in Fig.~\ref{CPhaseOldNew}, for each two-qubit sequence is undone by a corresponding single-qubit pulse of duration $\bar t$ applied to the spin pair $(\bullet\bullet)_a$.  

In order to understand how the operations shown in Fig.~\ref{u34} act on the $ab=11$ Hilbert space spanned by the states (\ref{5spins}), we introduce a pseudospin
\begin{eqnarray}
	\uparrow_f = (\blacktriangle(\bullet\blacktriangle)_{1/2})_{f}, \quad \downarrow_f = (\blacktriangle(\bullet\blacktriangle)_{3/2})_{f},
	\label{pseudospin}
\end{eqnarray}
for both the $f=\frac12$ and $\frac32$ sectors.  The actions of $U_3$, $\overline U_3$ and $U_4$ on these pseudospin spaces are worked out in Appendix \ref{matrixRepresentations}.  Consulting Eqs.~(\ref{f12}) and (\ref{f32}), the matrix representation of $U_4$ on these $f=\frac12$ and $f=\frac32$ pseudospins $\{\uparrow_f, \downarrow_f\}$ is
\begin{eqnarray}
	U^{f=1/2}_{4}(\phi) &=& e^{-i \phi/2}e^{i \phi \axis n_2\cdot \boldsymbol\sigma/2},
	\label{f12MT} \\
	U^{f=3/2}_{4}(\phi) &=& e^{-i\phi} \mathbb{1}.
	\label{f32MT}
\end{eqnarray}
For $f=\frac12$, $U_4(\phi)$ is a pseudospin rotation through angle $\phi$ about the axis $\axis n_2=(2\sqrt{2}/3, 0, -1/3)$ (see Note \cite{gauge_freedom}), which is indicated in Fig.~\ref{u34}(a).  Similarly, consulting Eqs.~(\ref{u3barApp}) and (\ref{u3App}) we have
\begin{eqnarray}
	\overline U^{}_{3}(\phi) &=& e^{-i\phi/2} e^{i\phi \axis z\cdot \boldsymbol\sigma /2},
	\label{u3barPseudo}\\
	U_{3}^{}(\phi) &=& e^{-i\phi/2} e^{i\phi \axis n_3\cdot \boldsymbol\sigma /2},
	\label{u3Pseudo} 
\end{eqnarray}
each of which results in the same pseudospin rotation through angle $\phi$ in both $f=\frac12$ and $\frac32$ sectors.  The rotation axes $\axis z$ and $\axis n_3 = (\frac{-4\sqrt 2}{9},0,-\frac79)$ are indicated in Figs.~\ref{u34}(b) and (c), respectively.  

For the operations $\overline U_5$ and $U_5$ to result in a \textsc{cphase} gate, Eq.~(\ref{cphase}) states that the $ab=11$ two-qubit states $((\bullet \blacktriangle)_{1/2}(\bullet\blacktriangle)_{1/2})_g$ with $g=0$ and 1, which can be obtained by combining the six-spin states shown in Fig.~\ref{qubits}(b) with Eq.~(\ref{introduceBlackTriangle}), need to be multiplied by $e^{i\phi}$.  Expanding these two-qubit states in the basis $(\bullet(\blacktriangle(\bullet\blacktriangle)_{1/2})_f)_g$ yields finite overlap with both $f=\frac12$ and $\frac32$ states [such concrete expansions are given by Eqs.~(23) and (24) in Ref.~\cite{zeuch14}].  $\overline U_5$ and $U_5$ must therefore multiply both pseudospin states $\uparrow_{1/2}$ and $\uparrow_{3/2}$ by the same phase factor $e^{-i\phi}$.  

Note that the similarity transformations due to $P$ and $Q$ in Eqs.~(\ref{u5bar}) and (\ref{u5}), respectively, have no effect on the $f=\frac32$ pseudospin sector because here $U_4$ is proportional to the identity [see Eq.~(\ref{f32MT})], so that
\begin{eqnarray}
	\overline U^{f=3/2}_{5}(\phi) = U^{f=3/2}_{5}(\phi) = e^{-i\phi} \mathbb{1}.
	\label{u5_32}
\end{eqnarray}
$\overline U_5$ and $U_5$ thus multiply the state $\uparrow_{3/2}$ by $e^{-i\phi}$ for arbitrary $P$ and $Q$.  The transformations (\ref{u5bar}) and (\ref{u5}) must then ensure that  $\uparrow_{1/2}$ is multiplied by the same phase factor.  This is accomplished by mapping $\axis n_2$, the rotation vector of $U_4$, to $-\axis z$, so that in the pseudospin basis $\{\uparrow_{1/2}, \downarrow_{1/2}\}$ we have
\begin{eqnarray}
	\overline U^{f=1/2}_{5}(\phi) = U^{f=1/2}_{5}(\phi) &=& e^{-i\phi/2} e^{i\phi (-\axis z)\cdot \boldsymbol\sigma /2} \nonumber \\
		&=& \text{diag}(e^{-i\phi}, 1).
\end{eqnarray}

The similarity transformation (\ref{u5bar}) found analytically in Ref.~\cite{zeuch14}, carried out by $P = U_4(\chi_1) \overline U_3(\phi_1)$, is visualized by the two intersecting cones in the top panel of Fig.~\ref{CPhaseOldNew}.  Here the green cone is described by rotating the vector $\axis n_2$ about the $z$ axis, and the yellow cone by rotating the vector $-\axis z$ about the $n_2$ axis.  The transformation due to $P$ then consists of a rotation of the vector $\axis n_2$ about the $z$-axis through angle $\phi_1$ to the intersection of the two cones, followed by a rotation about the $n_2$ axis through $\chi_1=2\pi - \phi_1$ to the negative $z$ axis.  The rotation angle is \cite{zeuch14}
\begin{eqnarray}
	\phi_1 = \cos^{-1}\frac{\axis n_2\cdot \axis z}{\axis n_2\cdot \axis z-1} = \cos^{-1}(1/4).
	\label{phi1}
\end{eqnarray}

The similarity transformation (\ref{u5}) for $U_5$, carried out by $Q = \overline U_3(\phi_3)U_3(\phi_2)$, is visualized by the green and gray cones shown in the lower panel of Fig.~\ref{CPhaseOldNew}.  Here the green cone is the same as that used before, while the gray cone is described by rotating $-\axis z$ about the $n_3$-axis.  The transformation due to $Q$ is then a rotation of $\bf {\hat n_2}$ about $\axis z$ through angle $\phi_2$ to the intersection of the cones, followed by an $n_3$-axis rotation through $\phi_3$ to $-\axis z$.  It is a simple exercise to show that the rotation angles are
\begin{eqnarray}
	\phi_2 &=& \cos^{-1}\left(- \frac{\axis n_{3}\cdot \axis z (1+\axis n_{2}\cdot\axis z)}{(\axis n_{2}\cdot\axis x) (\axis n_{3}\cdot\axis x)} \right) = \cos^{-1}(-7/8),\ 
	\label{phi2} \\
	\phi_3 &=& \cos^{-1}\left(- \frac{ {\bf \hat n_2}\cdot {\bf \hat z} + ({\bf \hat n_3}\cdot {\bf \hat z})^2}{({\bf \hat n_3}\cdot {\bf \hat x})^2} \right) = \cos^{-1}(-11/16).\ \nonumber \\
	\label{phi3}
\end{eqnarray}
These rotation angles then allow us to find the durations of the individual pulses making up $\overline U_3(\phi_2)$ and $U_3(\phi_3)$ [together with their inverses $\overline U_3(\chi_2)$ and $U_3(\chi_3)$] by solving Eqs.~(\ref{ttbar}) and (\ref{phi}).  The resulting full \textsc{cphase} pulse sequence for $U_5$ is given in Sec.~\ref{explicit} below.

In summary, the main difference between the two-qubit sequences of Ref.~\cite{zeuch14} and those constructed here, respectively shown in the top and bottom panels of Fig.~\ref{CPhaseOldNew}, is that the former are built of only two operations, $U_4$ and $\overline U_3$, while the latter are built of three, $U_4$, $\overline U_3$ and $U_3$.  These three operations result in pseudospin rotations about distinct axes so the latter construction is slightly more complex.  However, the resulting two-qubit sequences are more efficient, since the operation $U_3$ consists of fewer exchange pulses than $U_4$ (cf.~Fig.~\ref{CPhaseOldNew}).  

\begin{figure*}
	\includegraphics[width=\textwidth]{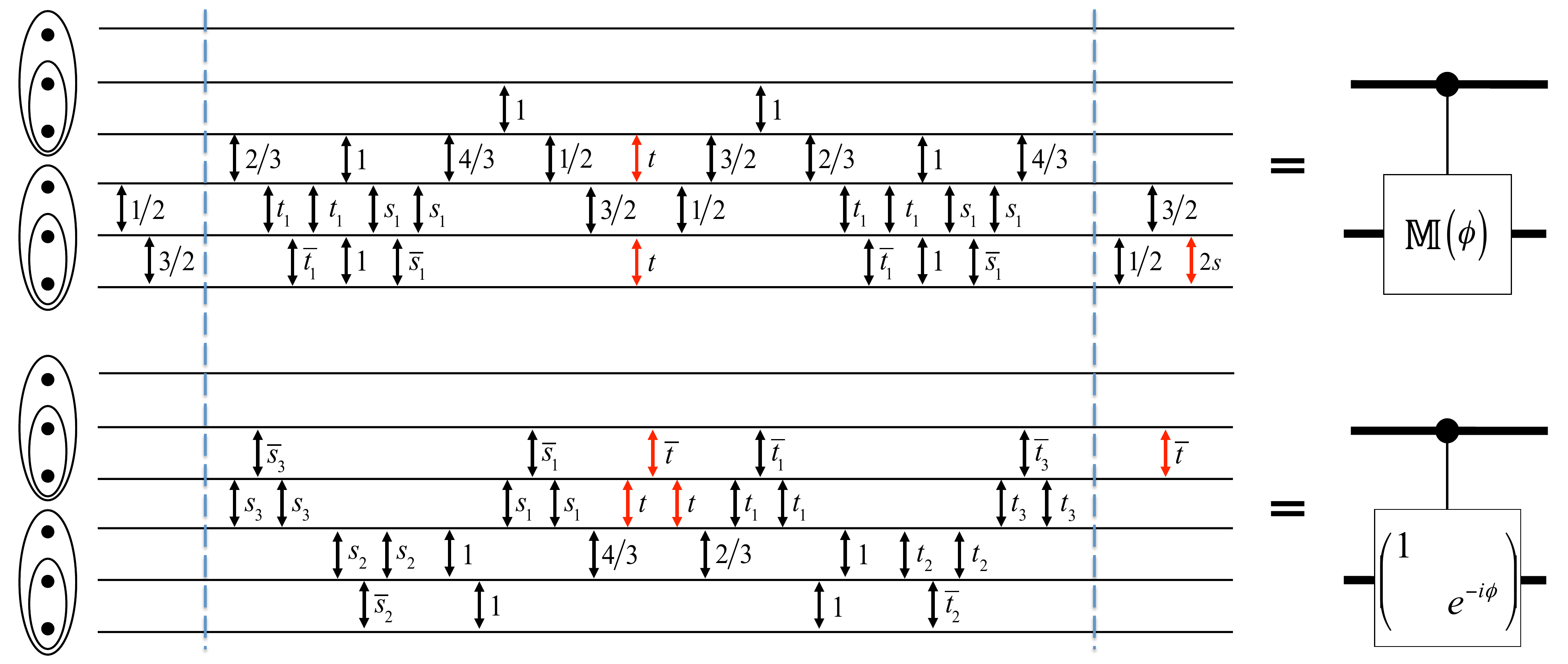}
	\caption{(Color online)  Example two-qubit gates applied to encoded three-spin qubits.  Top panel:  sequence shown in Fig.~\ref{observation1}(c) enacting a controlled-rotation gate (\ref{2qubit}) with $\mathbb M = \mathbb M(\phi)$.  Bottom panel:  sequence for $U_5$ shown in Fig.~\ref{CPhaseOldNew} (bottom panel) enacting a \textsc{cphase} gate.  The pulse durations independent of $\phi$, besides those given explicitly in the figure, are:  $t_1=0.426548$ \cite{zeuch14}, $t_2=0.469699$ and $t_3=0.685037$ (see main text); $\bar t_i$ for $i = 1, 2, 3$ obtained using Eqs.~(\ref{ttbar}); $s_i$ and $\bar s_i$ obtained using the relations $\bar t_i + \bar s_i = t_i+s_i = 2$ for $i=1,2,3$.  The durations of the pulses shown in red depend on the choice of $\phi$.  For the upper panel, $t$ and $s$ are obtained using $\phi(t)$ as given in Eq.~(\ref{phi(t)}); for the lower panel $t$ and $\bar t$ are obtained using Eqs.~(\ref{ttbar}) and (\ref{phi}).  The dashed lines on either side separate the core sequences from pulses that can be absorbed by single-qubit rotations.}
	\label{fullSequences}
\end{figure*}

\section{Explicit Pulse Sequences}
\label{explicit}

Figure \ref{fullSequences} shows explicit single-pulse representations of two-qubit gate sequences for the constructions presented above in Secs.~\ref{22} and \ref{23}, where single-qubit rotations are carried out by pulses acting before and after the core sequence as indicated by the dashed lines.  The gate shown on the top panel of Fig.~\ref{fullSequences} is locally equivalent to the arbitrary \textsc{cphase} gate shown in the bottom panel, both of which are characterized by $\phi \in [0, 2\pi]$.  Most pulses appearing in these sequences are independent of $\phi$, while those depending on $\phi$ are shown in red.  

The top panel of Fig.~\ref{fullSequences} shows an explicit two-qubit pulse sequence based on the schematic sequence of $T$, $S$ and \textsc{swap} operations shown in Fig.~\ref{observation1}(c).  As discussed in Sec.~\ref{22}, this sequence results in an arbitrary controlled-rotation gate of the form (\ref{2qubit}).  To obtain this sequence, we substituted for $T$ and $S$ the sequences derived in Appendix \ref{TS} and shown in Figs.~\ref{Tfig} and \ref{Sfig}, respectively.  In addition, simplifications yielding the single-qubit rotations discussed in the appendix have been applied.  For the parameterization of the controlled operation (\ref{parameterizeM}), i.e., $\mathbb M(\phi) = e^{i\xi(t)}e^{i \phi(t) \hat{\bf n}(t)\cdot\boldsymbol\sigma/2}$, we take from Eq.~(\ref{resultM}) that $\xi(t) = -\pi t/2$ and 
\begin{equation}
	\phi(t) = 2 \arccos((5\cos(\pi t/2)+3\cos(3\pi t/2))/8).
	\label{phi(t)}
\end{equation}
Given that $\phi(0) = 0$ and  $\phi(t_1) = 2\pi$ with $t_1 = 4 \arctan(\sqrt{2-\sqrt{3}})$, this pulse sequence can be used to carry out arbitrary controlled-rotation gates using values of $t\in[0, t_1]$.  

The lower panel of Fig.~\ref{fullSequences} shows an explicit pulse sequence based on the sequence for $U_5$ given in the lower panel of Fig.~\ref{CPhaseOldNew}.  This sequence is obtained by replacing $U_3$, $\overline U_3$ and $U_4$ by their single-pulse representations given in Fig.~\ref{34}.  To obtain the pulse durations $t_2$ and $t_3$ one needs to numerically solve Eqs.~(\ref{ttbar}) and (\ref{phi}) with $\phi=\phi_2$ and $\phi_3$, respectively, as given in Eqs.~(\ref{phi2}) and (\ref{phi3}).  As discussed in Sec.~\ref{23}, the two-qubit gate carried out by this sequence is an arbitrary \textsc{cphase} gate of the form (\ref{cphase}), where the values of $t$ and $\bar t$ depend on the choice of the phase $\phi$.

\section{Conclusions}
\label{conclusions}

For spin-based quantum computation in which quantum gates are carried out by exchange-pulse sequences, we have presented two different analytic constructions of entangling two-qubit gates.  Up to single-qubit rotations, the resulting sequences can be used to carry out arbitrary \textsc{cphase} gates of the form (\ref{cphase}), where the phase $\phi$ can be chosen freely by adjusting the durations of a small number of exchange pulses.  Other known two-qubit gate sequences either (i) result in a \textsc{cnot} gate \cite{divincenzo00, fong11, hsieh03, van19}, i.e., a gate locally equivalent to \textsc{cphase} with $\phi = \pi$, or (ii) consist of significantly more exchange pulses \cite{zeuch14}.  

When comparing the lengths of different two-qubit gate sequences, the minimum number of pulses depends on the interspin connectivity, which is determined by the spin layout assuming only nearest-neighbor pulses. Consider the specific example sequences given in Fig.~\ref{fullSequences} where the six spins making up two logical qubits are arranged in a linear array.  For these two sequences the total number of pulses are 28 (Fig.~\ref{fullSequences} top) and 25 (Fig.~\ref{fullSequences} bottom), where, for a basis-independent comparison, we have ignored single-qubit rotations, and continue to do so below.  We also consider the layout of maximal connectivity, in which the exchange interaction can be tuned directly between arbitrary spin pairs.  As can be deduced using the manipulations introduced in Appendix \ref{rearrange}, the number of required pulses for our sequences then reduces to the number of pulses different from \textsc{swap}; i.e., for each of the above two cases the total number of pulses drops from 28 to 22, and from 25 to 23.  Our sequences are thus significantly shorter than those of Ref.~\cite{zeuch14}, which consist of 39 pulses for either layout.  

The number of pulses required to carry out the Fong-Wandzura sequence is 18 for spins arranged on a linear array \cite{fong11}, while it is 12 for the case of highest connectivity (also easily established using the manipulations of Appendix \ref{rearrange}).  One way to obtain a \textsc{cphase} gate (\ref{cphase}) with $\phi \neq \pi$ would be to sandwich a single-qubit operation between two \textsc{cnot}s; using the optimal Fong-Wandzura sequence this requires at least $2\times 18+1 = 37$ or $2\times 12+1 = 25$ pulses.  For either case, our sequences are thus more efficient (see, however, Note \cite{FongPrivate}).  

To summarize, in Sec.~\ref{22} we generalize the construction of the Fong-Wandzura sequence \cite{fong11}, which can be used to enact a \textsc{cnot} gate, to a new construction for controlled-rotation gates locally equivalent to arbitrary \textsc{cphase}.  Starting from the original Fong-Wandzura sequence, we did not simply change the durations of individual pulses but rather altered its fundamental structure.  The second two-qubit gate construction, presented in Sec.~\ref{23}, makes use of smaller pulse sequences whose operations conserve the total spins of certain spin pairs they act on \cite{zeuch14}.  By doing this, a large subspace of the complete Hilbert space associated with the six spins encoding two logical qubits is rendered trivial.  The resulting family of two-qubit sequences is, in essence, a streamlined version of that derived in Ref.~\cite{zeuch14}.  In both of these cases, ideas developed in Refs.~\cite{zeuch16} and \cite{zeuch14} have been generalized and used to construct entirely new pulse sequences for exchange-only quantum computation.

\begin{acknowledgments}

We thank Joel Pommerening, Veit Langrock, Jorge Piecarewicz, Martin Savage, Pia D\"oring and David DiVincenzo for useful discussions.  The National High Magnetic Field Laboratory is supported by the National Science Foundation through NSF/DMR-1644779 and by the State of Florida.

\end{acknowledgments}

\appendix

\section{Rearranging the Fong-Wandzura Pulse Sequence}
\label{rearrange}

\begin{figure}
		\includegraphics[width=\columnwidth]{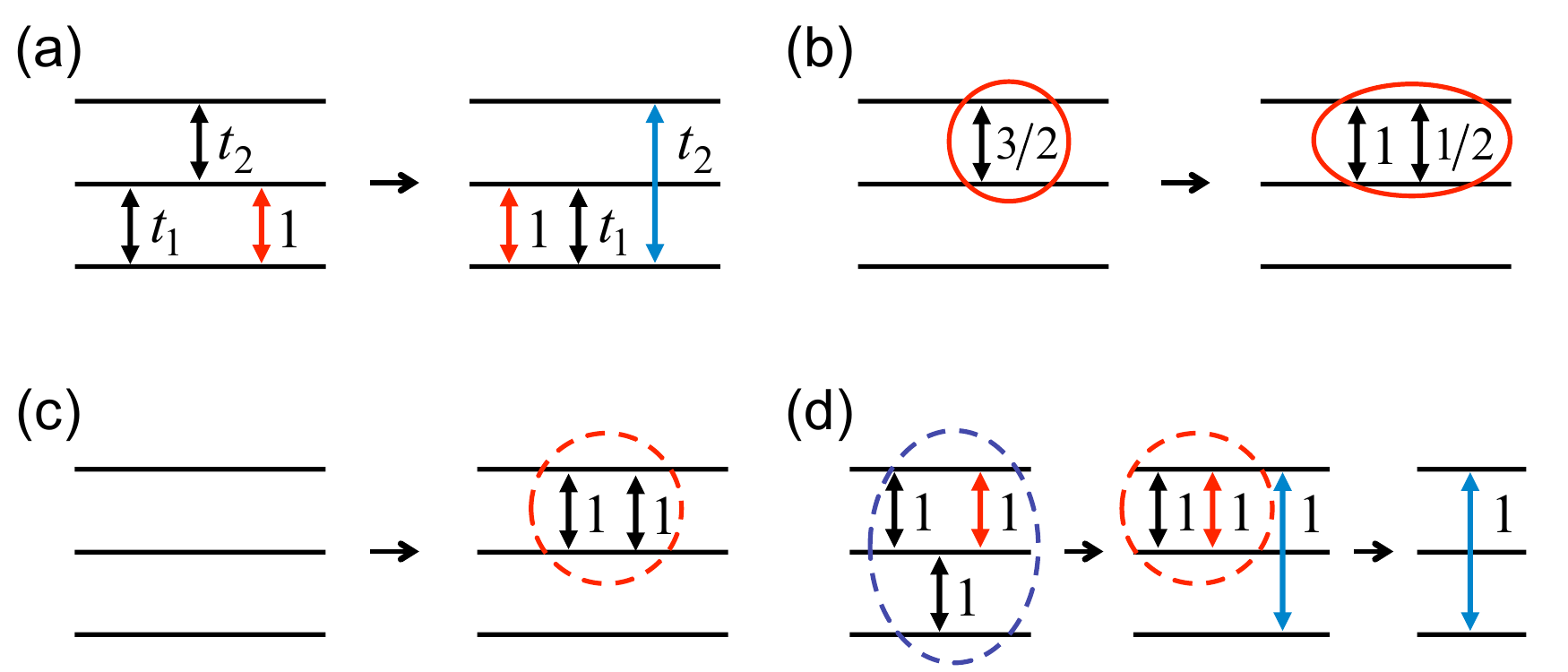}
	\caption{(Color online)  Examples of elementary pulse-sequence manipulations. \textsc{swap}s moved past other exchange pulses are shown in red, and pulses whose connectivity is altered by such a move are shown in blue.  (a)  Moving a \textsc{swap} past two pulses of generic durations $t_1$ and $t_2$.  (b)  Rewriting an inverse $\sqrt{\textsc{swap}}$ as a \textsc{swap} followed by a $\sqrt{\textsc{swap}}$---affected pulses are enclosed by red ovals. (c)  Inserting a pair of \textsc{swap}s---new pulses are enclosed by the red dashed circle.  (d) Simplification of a three-\textsc{swap} sequence using (a) followed by the reverse of (c).  The three nearest-neighbor \textsc{swap}s encircled by a blue dashed oval are thus equivalent to a single next-nearest neighbor \textsc{swap}.  Note that all reversed diagrams are true as well.}
	\label{examples}
\end{figure}

In this appendix we show explicitly that the Fong-Wandzura pulse sequence, as it is published in Ref.~\cite{fong11}, is equivalent to the two-qubit gate sequence derived in Ref.~\cite{zeuch16}.  Both of these sequences, which are shown in Fig.~\ref{FWsequences}, consist only of nearest-neighbor exchange pulses, assuming the spins that these sequences act on are placed on a linear array.  These sequences can be transformed into one another by carrying out a series of elementary manipulations in which pairs of \textsc{swap}s, i.e., exchange pulses of duration 1, are inserted into a pulse sequence, or single $\textsc{swap}$s are moved past, or combined with, other pulses.  

\begin{figure*}
	\includegraphics[width=\textwidth]{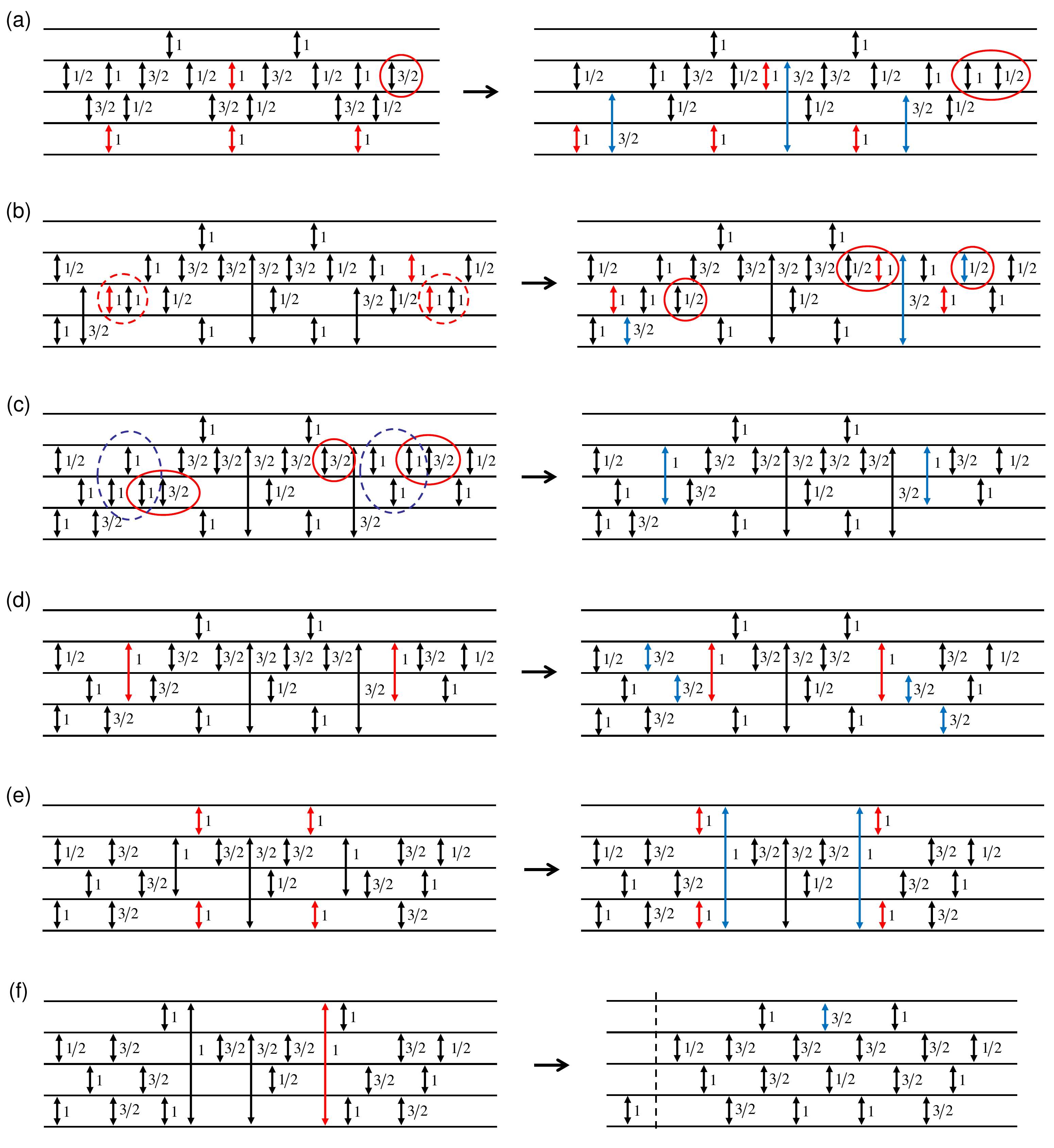}
	\caption{(Color online)  Series of elementary manipulations used to turn the core Fong-Wandzura sequence analytically derived in Ref.~\cite{zeuch16} into its equivalent form published originally in Ref.~\cite{fong11}, thus proving the equality stated in Fig.~\ref{FWsequences}.  In part (f) a \textsc{swap}, which can be absorbed into a single-qubit rotation, is separated from the original Fong-Wandzura sequence.}
	\label{rearrangement}
\end{figure*}

Figure \ref{examples} exemplifies a number of such manipulations by means of short pulse sequences acted on three spin-$\frac12$ particles.  Figure \ref{examples}(a) shows two equivalent sequences.  The one on the left-hand side (LHS) of the figure consists of three nearest-neighbor pulses, beginning with a pulse of duration $t_1$ acting on the lower two spins, followed by a pulse of duration $t_2$ acting on the upper two spins, and ending with a \textsc{swap} acting on the lower two spins.  The equivalent sequence on the right-hand side (RHS) of the same figure consists of a \textsc{swap} followed by a $t_1$-pulse, both of which act on the lower two spins, and ends with a next-nearest neighbor $t_2$-pulse acting on the uppermost and lowermost spins.  The equivalence of these sequences becomes evident upon replacing the \textsc{swap}s with spin permutations.  One way to manipulate pulse sequences is thus to move \textsc{swap}s past other pulses while taking note if after such a move these other pulses act on different spins than before.  As also shown in Fig.~\ref{examples}(a), in this appendix we adopt the convention that \textsc{swap}s moved past other pulses are shown in red, and exchange pulses altered by such a move are shown in blue.  

Another way of manipulating a pulse sequence, which is exemplified in Fig.~\ref{examples}(b), is that of replacing an inverse $\sqrt{\textsc{swap}}$ (pulse of duration 3/2) with a \textsc{swap} followed by a $\sqrt{\textsc{swap}}$ (pulse of duration 1/2).  This is allowed because an exchange pulse acting on a pair spins of duration $t_1+t_2$ is equivalent to two consecutive pulses acting on the same pair of spins with durations $t_1$ and $t_2$.  As a last example of an elementary manipulation, consider Fig.~\ref{examples}(c).  As shown in the figure, we can insert (remove) a pair of \textsc{swap}s into (from) a pulse sequence because the effect of two consecutive \textsc{swap}s is the same as applying the identity.  Note that when the number of pulses changes due to a manipulation step [as in Figs.~\ref{examples}(b) and (c)] we enclose the involved pulses inside an oval.  

As an example of using some of the above manipulations, consider the three pulse sequences shown in Fig.~\ref{examples}(d).  The leftmost sequence consists of three nearest-neighbor \textsc{swap}s (enclosed by a blue oval) with the first and the last acting on the upper two spins, and the central \textsc{swap} acting on the lower two spins.  In the first step we move the rightmost \textsc{swap}, shown in red, past the central pulse, yielding the second sequence in Fig.~\ref{examples}(d).  Due to this move, the now rightmost pulse, shown in blue, is a next-nearest neighbor \textsc{swap} that acts on the uppermost spin and the lowermost spin.  Finally, since after this move the leftmost \textsc{swap} and the red \textsc{swap} are located directly next to each other we are allowed to remove this pair of pulses from the sequence.  

With these basic manipulation steps in hand we are now ready to convert the two-qubit gate sequence shown on the LHS of Fig.~\ref{FWsequences}, which was derived in Ref.~\cite{zeuch16}, into the originally published Fong-Wandzura sequence \cite{fong11} shown on the RHS of the same figure.  Figure \ref{rearrangement} contains the transformation that shows this equivalence, where we apply the two-qubit gate sequences on five of the six spins that are used to encode two qubits.  

Beginning with the sequence shown on the LHS of Fig.~\ref{FWsequences}, in Fig.~\ref{rearrangement}(a) we move the four \textsc{swap}s shown in red to the left; their original and final positions are given on the LHS and RHS of the figure, respectively.  Similar to the example of Fig.~\ref{examples}(a), the pulses that are shown in blue have been altered by these moves.  Furthermore, similar to Fig.~\ref{examples}(b), the rightmost inverse $\sqrt{\textsc{swap}}$ in the sequence on the LHS of Fig.~\ref{rearrangement}(a) has been replaced with a \textsc{swap} and a $\sqrt{\textsc{swap}}$ (as indicated by the red ovals).  The next step begins with the sequence shown on the LHS of Fig.~\ref{rearrangement}(b), which is the result of taking the previous sequence [i.e. that shown on the RHS of Fig.~\ref{rearrangement}(a)] and, similar to the example in Fig.~\ref{examples}(c), inserting two pairs of \textsc{swap}s, which are enclosed in dashed circles.  The three red \textsc{swap}s on the LHS of Fig.~\ref{rearrangement}(b) are then moved towards the left with their final positions given in the sequence on the RHS of the figure.  

When turning to the sequence on the LHS of Fig.~\ref{rearrangement}(c), the $\sqrt{\textsc{swap}}$ and its neighboring \textsc{swap} enclosed by an oval in the sequence on the RHS of Fig.~\ref{rearrangement}(b) are replaced with an inverse $\sqrt{\textsc{swap}}$.  Furthermore, the two individually circled $\sqrt{\textsc{swap}}$s on the RHS of Fig.~\ref{rearrangement}(b) have each been replaced by a pair of pulses in Fig.~\ref{rearrangement}(c).  In the transition to the RHS of this figure, we use the identity shown in Fig.~\ref{examples}(d) to replace each of the three \textsc{swap}s encircled by a blue dashed oval by a single next-nearest neighbor \textsc{swap}.

The remaining task is to remove these two next-nearest neighbor \textsc{swap}s by placing them directly next to each other.  For clarity, this is done in several steps shown in Figs.~\ref{rearrangement}(d)-(f).  First, in (d) we move both of these red \textsc{swap}s towards the center of the sequence where, as usual, pulses that have been altered are shown in blue.  Rather than moving the same pulses further towards each other, in (e) we take the four red nearest-neighbor \textsc{swap}s and move each of them one step towards the outside of the sequence, thus altering the next-nearest neighbor \textsc{swap}s.  Finally, in (f) we take the \textsc{swap} shown in red and move it towards the left where it is combined with its equivalent \textsc{swap}, and thus removed.  Note that in this last manipulation only a single exchange pulse is altered.  

In the resulting sequence of nearest-neighbor pulses, which is shown on the RHS of Fig.~\ref{rearrangement}(f), we use a dashed line to separate out a \textsc{swap}, which can be absorbed by a single-qubit rotation.  This final sequence is thus locally equivalent to the that given on the RHS of Fig.~\ref{FWsequences}.

% Appendix B
\section{Explicit Pulse Sequences for the $T$ and $S$ Operations}
\label{TS}

In Sec.~\ref{22}, a set of two-qubit gate sequences is constructed using the operations $T$ and $S$ defined in Eqs.~(\ref{T}) and (\ref{S}), respectively.  In this appendix we design example pulse sequences that can be used to carry out these operations, and which are the ones used for the explicit sequences presented in Sec.~\ref{explicit}.  

As shown in Fig.~\ref{observation1}(c), the operations $T$ and $S$ are applied to four spin-$\frac12$ particles, three of which, initialized with total spin $\frac12$, are represented by the symbol $\bigstar$.  In order to design explicit pulse sequences, we first abandon the notation $\bigstar$ by effectively reversing Eq.~(\ref{bigstar}), so that the Hilbert space of $\bullet$ and $\bigstar$ is now spanned by four states
\begin{eqnarray}
	(\bullet\bigstar)_d \rightarrow (\bullet(\bullet(\bullet\bullet)_b)_{c=1/2})_d
	\label{unpacked}
\end{eqnarray}
with $b=0$ or 1 for both $d = 0$ and 1.  Since an exchange pulse acting on the two leftmost spins on the RHS of Eq.~(\ref{unpacked}) does not conserve the total spin of the rightmost three spins, denoted by $c$, we need to consider a five-dimensional Hilbert space spanned by the states $(\bullet(\bullet(\bullet\bullet)_b)_{c})_d$ with $bc=0\frac12$ and $1\frac12$ for $d=0$ and $bc=0\frac12$, $1\frac12$ and $1\frac32$ for $d=1$.  Recall, however, that in Sec.~\ref{22} we imposed the condition that applying the full sequence of $T$ or $S$ conserve this quantum number $c$.  

\begin{figure}
	\includegraphics[width=\columnwidth]{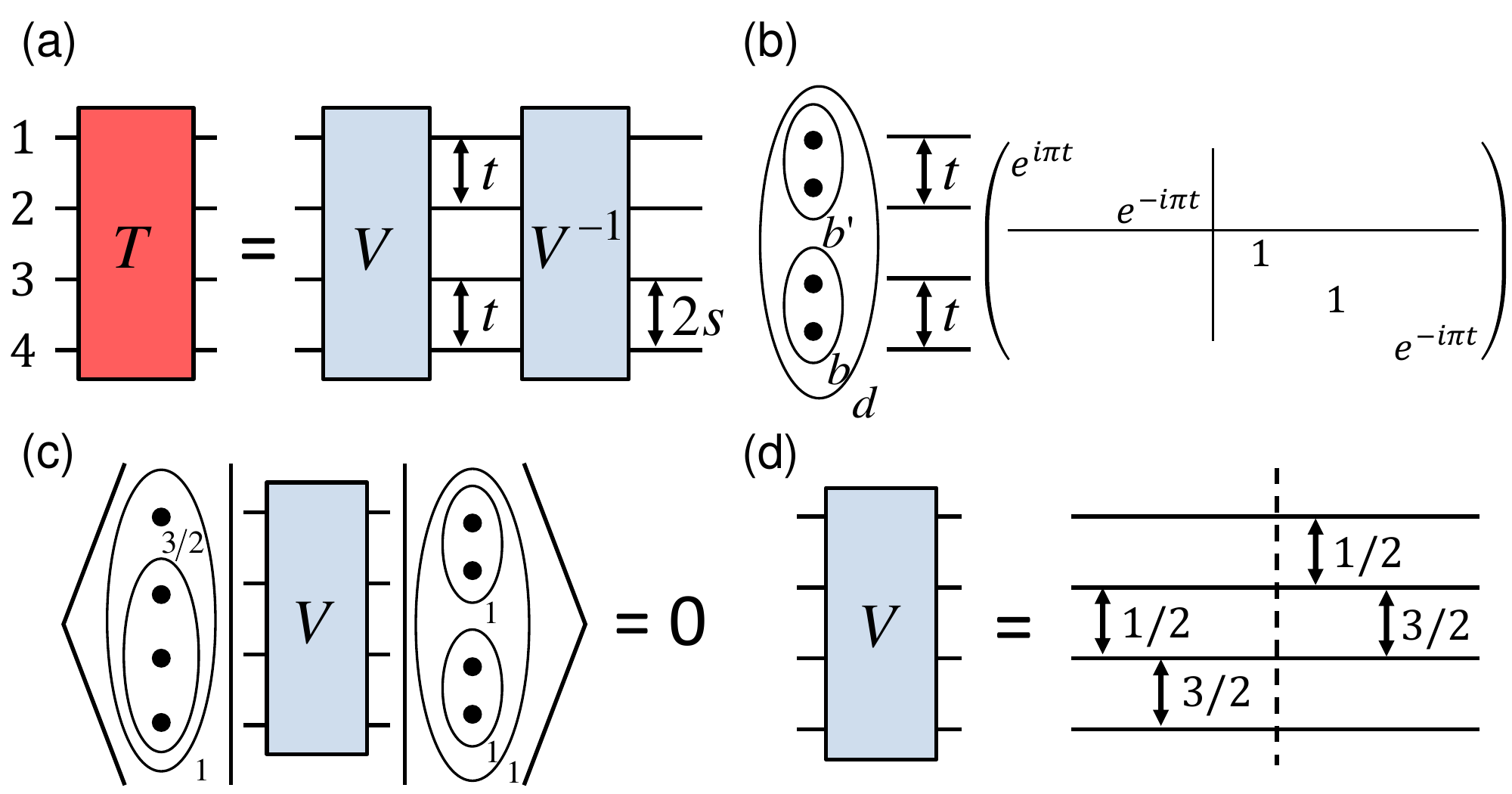}
	\caption{Construction of a pulse sequence for the $T$ operation.  (a) Schematic pulse sequence for $T$ applied to four spins labeled 1 through 4, where $s = 2 -  t$.  (b) Matrix representation of the operation due to the central $t$-pulses in the indicated basis with state ordering $bb'd = \{000, 110 | 011, 101, 111\}$; in the matrix solid lines separate different total-spin sectors from one another.  (c) Crucial constraint placed on the operation $V$.  (d) Explicit pulse sequence for $V$.  The dashed line separates the optimal $V$-sequence satisfying (c), $V_0 = U_{23}(\frac32)U_{12}(\frac12)$, from the additional pulses used to satisfy the condition imposed by Eq.~(\ref{T}).}
	\label{Tfig}
\end{figure}

Figure \ref{Tfig} contains the essence of our derivation of a pulse sequence that realizes $T$.  Figure \ref{Tfig}(a) shows the sequence used for $T$ as it is applied to four spins labeled 1 through 4, that is,
\begin{eqnarray}
	T = U_{34}(2s) V^{-1} U_{12}( t)U_{34}(t) V.
	\label{T1}
\end{eqnarray}
Here $U_{ij}$, as introduced in Eq.~(\ref{pulse}), represents an exchange pulse acting on individual spins $i$ and $j$, and $V$ represents a pulse sequence yet to be determined.  The sequence (\ref{T1}) can then be understood as a similarity transformation of the two central $t$-pulses $U_{12}(t)U_{34}(t)$ carried out by $V$, which is followed by the operation $U_{34}(2s)$.  

To find the matrix representation of this $T$ operation, first note that using Eq.~(\ref{pulse}) it is straightforward to find the matrix associated with the two central pulses of duration $t$ in the basis $((\bullet\bullet)_{b'}(\bullet\bullet)_{b})_d$.  The resulting matrix, for simplicity given up to an overall phase factor, is shown in Fig.~\ref{Tfig}(b).  We now examine the effect of the remaining operations in Eq.~(\ref{T1}) on this central operation in each of the total-spin $d=0$ and 1 sectors.  

First considering the total-spin $d=1$ subspace, we place the constraint shown in Fig.~\ref{Tfig}(c) on the operation $V$.  This constraint implies that after applying $V$ to the state $(\bullet(\bullet\bullet\bullet)_{c=3/2})_{1}$, the outcome has no overlap with the state $((\bullet\bullet)_{b'=1}(\bullet\bullet)_{b=1})_1$, and accordingly lies completely in the $((\bullet\bullet)_{b'}(\bullet\bullet)_b)_1$ sector with $bb'=01$ and 10.  Note that in this subspace the operation $U_{12}(t)U_{34}(t)$ is proportional to the identity [cf.~Fig.~\ref{Tfig}(b)] and thus leaves the state unchanged, so that the operation $V^{-1}$ maps this state back to the original state, $(\bullet(\bullet\bullet\bullet)_{3/2})_{1}$.  Since the operation $V^{-1} U_{12}( t)U_{34}( t) V$ thus maps this $c=\frac32$ state back onto itself, and further the final pulse in Eq.~(\ref{T1}), $U_{34}(2s)$, merely applies a phase factor to this state, it follows that the $T$ operation also maps the two-dimensional $c=\frac12$ sector onto itself.  

In the derivation of the Fong-Wandzura sequence in Ref.~\cite{zeuch16} an argument similar to that just made is used to find a pulse sequence for the $R$-operation of Fig.~\ref{FWobservation1}(a).  In doing this, the optimal sequence for such a $V$ operation satisfying the constraint in Fig.~\ref{Tfig}(c) is analytically determined to be $V_0 = U_{23}(\frac32)U_{12}(\frac12)$.  

Figure \ref{Tfig}(d) shows the $V$ sequence used in our construction, which is $V=U_{23}(\frac32)U_{12}(\frac12)U_{34}(\frac32)U_{23}(\frac12) = V_0U_{34}(\frac32)U_{23}(\frac12)$.  We use this longer sequence for $V$ in order to obtain the matrix representation of $T$ given in Eq.~(\ref{T}) (see Note \cite{V0-is-enough}).  Given that $V_0$ satisfies the constraint in Fig.~\ref{Tfig}(c), it is straightforward to see that our longer $V$ sequence also satisfies that constraint.  This is most easily seen by replacing the $V$ sequence in Fig.~\ref{Tfig}(c) with its single-pulse representation shown in Fig.~\ref{Tfig}(d).  The two pulses to the left of the dashed line can then be absorbed by the state $(\bullet(\bullet\bullet\bullet)_{3/2})_1$ at the cost of applying simple overall phase factors (since any pair of spins within the oval with total spin $\frac32$ has total spin $1$), which do not alter the fact that this overlap vanishes.  

The matrix representation of $T$ in the $d=1$ sector is determined in Appendix \ref{four}.  This is done by finding the matrix of each pulse in an appropriate eigenbasis using Eq.~(\ref{pulse}), and then carrying out the required basis changes to the $b$-basis (\ref{unpacked}) for $d=1$;  these basis changes are summarized in Appendix \ref{basis_changes}.  In this $b$-basis with the state ordering $b = \{0, 1\}$ we take from Eq.~(\ref{AppendixM}) that
\begin{eqnarray}
	\mathbb{M} &=& e^{-i\pi t/2}e^{i\pi (2-t){\hat {\bf z}}\cdot \boldsymbol{\sigma}} e^{-i\pi t\hat{\bf n}_1\cdot \boldsymbol \sigma/2},
	\label{resultM}
\end{eqnarray}
where $\hat{\bf n}_1 = (\sqrt 3/4, -\sqrt 3/2, 1/4)$.  

To determine the unitary operation of $T$ on the $d=0$ Hilbert space, note that here the two four-spin states in the basis shown in Fig.~\ref{Tfig}(b) are given by $((\bullet\bullet)_{b'}(\bullet\bullet)_{b})_{0}$ with $bb'=00$ or 11, and thus $b'=b$.  A direct result of this is that
\begin{eqnarray}
	\quad U_{12}(t) = U_{34}(t), \qquad \qquad (d=0),
	\label{equalU's}
\end{eqnarray}
so that the two central pulses within $V$, $U_{12}(\frac12)$ and $U_{34}(\frac32)$, cancel one another.  Since the outer two pulses are inverses of each other as well, we conclude that $V = \mathbb{1}$ ($d=0$).  Equation (\ref{equalU's}) further implies that the remaining three pulses within the $T$ sequence [shown explicitly in Fig.~\ref{Tfig}(a)] cancel, so that $T = \mathbb{1}$ for $d=0$ as required by Eq.~(\ref{T}).  

Before we turn our attention to finding a sequence for the $S$ operation, we make a comment on the $T$ sequence of Fig.~\ref{Tfig}(a), which, upon unpacking $V$ using Fig.~\ref{Tfig}(d), can be written as
\begin{eqnarray}
	T &=& U_{34}(2s) U_{23}(\tfrac32) U_{34}(\tfrac12) \times \nonumber \\
		&&[ U_{12}(\tfrac32) U_{23}(\tfrac12)U_{12}(t)U_{34}(t) U_{12}(\tfrac12)U_{23}(\tfrac32)] \times \nonumber \\
		&& U_{34}(\tfrac32) U_{23}(\tfrac12).
	\label{TSeparate}
\end{eqnarray}
Recall that $T$ is the central operation in the two-qubit gate sequence shown in Fig.~\ref{observation1}(c).  As shown in the top panel of Fig.~\ref{fullSequences}, all five pulses outside the square brackets in Eq.~(\ref{TSeparate}), i.e., $U_{34}(2s) U_{23}(\tfrac32) U_{34}(\tfrac12)$ and $U_{34}(\tfrac32) U_{23}(\tfrac12)$, have been pulled out of the two-qubit sequence (thus playing the role of single-qubit rotations).  The reason we are allowed to do this is that each of these five pulses commutes with $S$, since they exclusively act on the internal Hilbert space of the lower logical qubit with state label $b$ [cf.~top panel of Fig.~\ref{fullSequences}], and the surrounding $S$ operations act on this lower-qubit space as the identity [cf.~Eq.~(\ref{S})].  

\begin{figure}
	\includegraphics[width=\columnwidth]{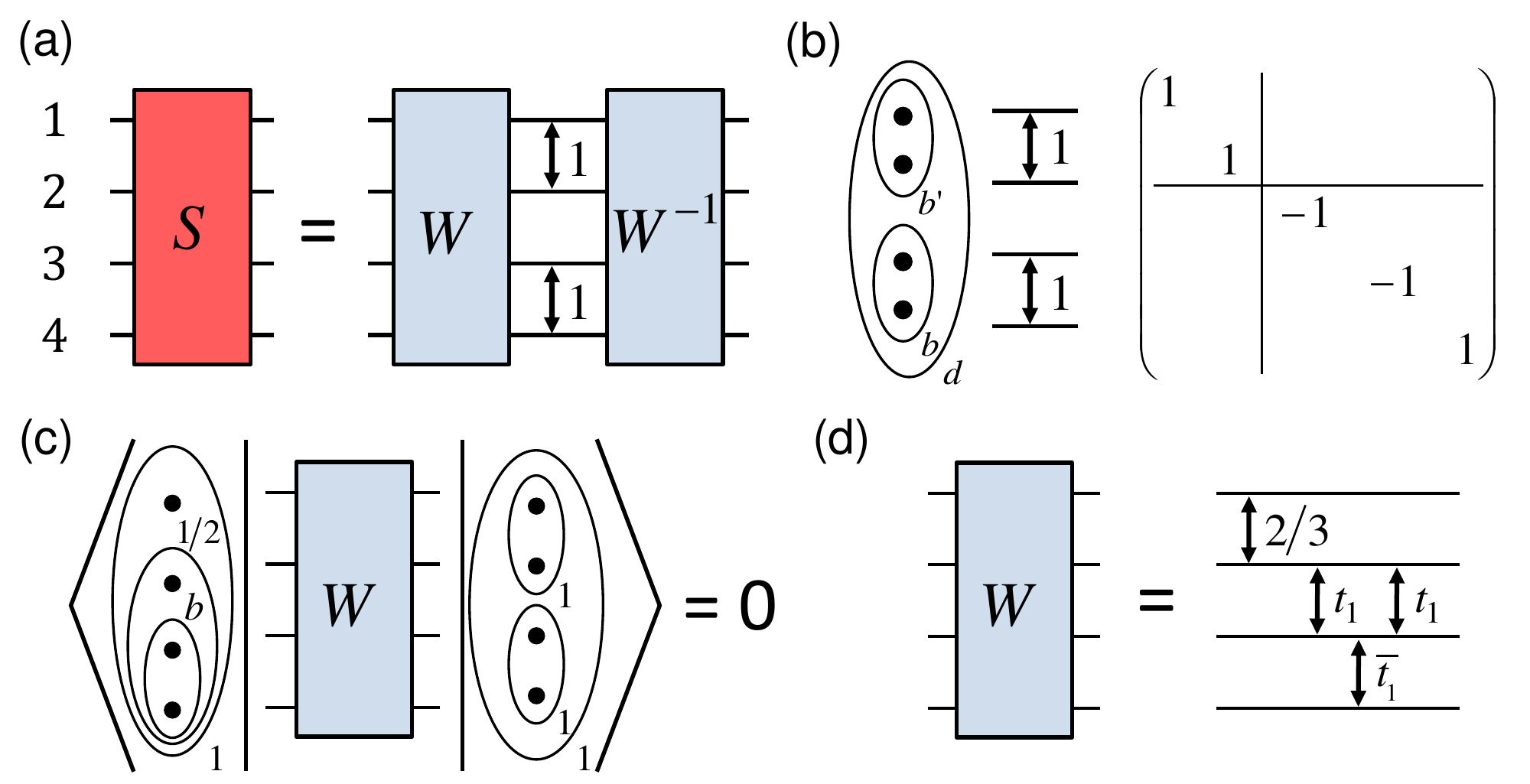}
	\caption{Construction of a pulse sequence for the elevated \textsc{swap} operation $S$.  (a)  Schematic sequence for $S$ applied to four spins labeled 1 through 4.  (b)  Matrix representation of the operation due to the central \textsc{swap} pulses in the indicated basis with state ordering $bb'd = \{000, 110 | 011, 101, 111\}$; similar to Fig.~\ref{Tfig}, solid lines separate different total-spin sectors from one another.  (c) Constraints placed on $W$ for both $b=0$ and 1.  (d)  Explicit sequence for $W$ with pulse durations $t_1$ and $\bar t_1$ given in Fig.~\ref{fullSequences}.}
	\label{Sfig}
\end{figure}

Figure \ref{Sfig}(a) shows a four-spin pulse sequence that realizes the $S$ operation, which, again labeling the spins 1 though 4, can be written as
\begin{equation}
	S = W^{-1} U_{12}(1) U_{34}(1)W.
	\label{S1}
\end{equation}
Similar to the schematic sequence $V$ in the sequence (\ref{T1}) for the $T$ operation, here $S$ is formulated using an operation $W$ whose pulse sequence is yet to be determined.  Note that $W$ carries out a similarity transformation on the central two \textsc{swap}s.  

We evaluate the $S$-sequence by first noting the matrix representation of the central \textsc{swap}s in Fig.~\ref{Sfig}(b) in the indicated $bb'$-basis $((\bullet\bullet)_{b'}(\bullet\bullet)_b)_{d}$, which is obtained easily using Eq.~(\ref{pulse}).  Comparing this matrix with that in Eq.~(\ref{S}) [given in the basis (\ref{unpacked})], we find agreement not only for $d=0$, but also for $d=1$ upon focusing on the $bb'=01$, 10 matrix sector in Fig.~\ref{Sfig}(b).  Since these matrices are given in different bases, the operation $W$ in Eq.~(\ref{S1}) needs to map the $b$-basis (\ref{unpacked}) to the $bb'=01$, 10 sector of the $bb'$ basis for both $d=0$ and 1.  

Note that since for $d=0$ the central \textsc{swap}s in Eq.~(\ref{S1}) carry out the identity operation [see Fig.~\ref{Sfig}(b)], here this change of bases is trivially accomplished by any sequence $W$.  In case of $d=1$, however, we need to ensure that applying $W$ to the states $(\bullet(\bullet(\bullet\bullet)_b)_{1/2})_{1}$ with $b=0$ or 1, which span the $b$-basis (\ref{unpacked}) for $d=1$, yield states orthogonal to $((\bullet\bullet)_{1}(\bullet\bullet)_{1})_{1}$.  As shown in Fig.~\ref{Sfig}(c), we place this very constraint on $W$.  Note that since the operation due to the central \textsc{swap}s in Eq.~(\ref{S1}) is proportional to the identity in the $bb'=10$, 01 sector [again, see Fig.~\ref{Sfig}(b)], the precise form of the basis change carried out by $W$ is irrelevant [i.e., the condition shown in Fig.~\ref{Sfig}(c) is sufficient].  The pulse sequence for $W$ shown in Fig.~\ref{Sfig}(d) has been designed in Ref.~\cite{zeuch14} to carry out such a basis change.

% Appendix C
\section{Computation of Pseudospin Rotations}
\label{pseudospin_rotations}

In Sec.~\ref{23}, two-qubit gate sequences are constructed using the  operations $U_3$, $\overline U_3$ and $U_4$.  The explicit sequences for these operations are given in Fig.~\ref{34} where $U_3$ and $\overline U_3$ are applied to three spin-$\frac12$ particles while $U_4$ is applied to four spin-$\frac12$ particles.  In Fig.~\ref{u34} these operations act on the five spins
\begin{eqnarray}
	((\bullet\bullet)_{a}(\bullet(\bullet\bullet)_{b})_c)_f,
	\label{5spinsApp}
\end{eqnarray}
where $a, b=0$ or 1 and $c, f=\frac12$ or $\frac32$.  

As described in Sec.~\ref{23}, we effectively reduce the Hilbert space dimensionality by setting $a, b\rightarrow 1$.  Upon introducing an effective spin-$1$ particle, $\blacktriangle = (\bullet\bullet)_1$, as in Eqs.~(\ref{introduceBlackTriangle}) and (\ref{5spins}), this Hilbert space is spanned by the states
\begin{eqnarray}
	(\blacktriangle(\bullet\blacktriangle)_c)_f
	\label{effApp}
\end{eqnarray}
with $cf=\frac12\frac12$, $\frac32\frac12$, $\frac12\frac32$ and $\frac32\frac32$, and so breaks into two two-dimensional sectors with total spin $f=\frac12$ and $\frac32$.  

In what follows, we discuss in \ref{simplification} why we are allowed to simplify the Hilbert space basis (\ref{5spinsApp}) to (\ref{effApp}), and what needs to be taken into account when doing so.  \ref{matrixRepresentations} then contains derivations of the matrix representations of the operations $U_3$, $\overline U_3$ and $U_4$.

\subsection{Effective Hilbert Space}
\label{simplification}

As discussed in Sec.~\ref{23}, the operations $U_3$, $\overline U_3$, and $U_4$ as arranged in Fig.~\ref{u34} conserve the quantum numbers $a$ and $b$.  It is thus natural to divide the Hilbert space into four different sectors with $ab = 00$, 01, 10 and 11, which can be considered independently.  We now explain why each of these three operations acts trivially on the states (\ref{5spinsApp}) with $a=0$ or $b=0$, a fact that allows us to focus on the effective Hilbert space spanned by the states (\ref{effApp}).  

The operation $U_4$ acting on the spins (\ref{5spinsApp}) as shown in Fig.~\ref{u34}(a) is accompanied on either side by a permutation of one spin-$\frac12$ particle with two spin-$\frac12$ particles, introduced as \textsc{powt} in Sec.~\ref{23},
\begin{eqnarray}
	(\bullet(\bullet\bullet)_b)_c \longleftrightarrow ((\bullet\bullet)_b\bullet)_c.
	\label{generalizedSWAP}
\end{eqnarray}
Because of this, the sequence of $U_4$ can equivalently be regarded as directly acting on the leftmost four spins in the basis
\begin{eqnarray}
	((\bullet\bullet)_{a}((\bullet\bullet)_{b}\bullet)_c)_f.
	\label{natural'}
\end{eqnarray}
If $a=0$ or $b=0$, respectively, these five spins are in the states
\begin{eqnarray}
	((\bullet\bullet)_{0}((\bullet\bullet)_{b}\bullet)_c)_f = ( ((\bullet\bullet)_{0}(\bullet\bullet)_{b})_{b} \bullet)_f
	\label{natural0}
\end{eqnarray}
or
\begin{eqnarray}
	((\bullet\bullet)_{a}((\bullet\bullet)_{0}\bullet)_c)_f = ( ((\bullet\bullet)_{a}(\bullet\bullet)_{0})_{a} \bullet)_f,
	\label{natural1}
\end{eqnarray}
where $c, f = \frac12$ or $\frac32$.  Note that for both of these states as expressed on the RHS of these equations, the leftmost four spins are given in the natural basis shown on the far left in Fig.~\ref{34}(c), in which we introduced the operation $U_4$.  It then follows from Eq.~(\ref{u4a=0}) that the action of $U_4$ on the state (\ref{5spinsApp}) with $a=0$ [taking into account the surrounding \textsc{swap}s (\ref{generalizedSWAP})] is that of applying a simple phase factor independent of the value of $b$.  Similarly, from Eq.~(\ref{u4}) we find that the action of $U_4$ on all $b=0$ states (\ref{5spinsApp}) is that of the identity.  

The operation $\overline U_3$ shown in Fig.~\ref{u34}(b) acts on the lowermost three spins, and is thus independent of $a$.  If $b=0$, the spins (\ref{5spinsApp}) are in the states
\begin{eqnarray}
	((\bullet\bullet)_a(\bullet(\bullet\bullet)_{0})_{1/2})_{f},
	\label{U3barStateB=0}
\end{eqnarray}
where $a=0$ or 1 and $f=\frac12$ or $\frac32$.  The action of $\overline U_3$ on the rightmost three spins is given by Eq.~(\ref{u3bar}) for the case of $a' = 0$, and is that of applying an overall phase factor to all $b=0$ states.  Similarly, the operation $U_3$ shown in Fig.~\ref{u34}(c) acts on the uppermost three spins and is therefore independent of $b$.  Since for $a=0$ the spins (\ref{5spinsApp}) are in the states
\begin{eqnarray}
	((\bullet\bullet)_{0}(\bullet(\bullet\bullet)_{b})_{c})_{f=c} = (((\bullet\bullet)_{0}\bullet)_{1/2}(\bullet\bullet)_{b})_{f=c},
	\label{U3barStateA=0}
\end{eqnarray}
where $b = 0$ or 1 and $c=\frac12$ or $\frac32$, the action of $U_3$ on the leftmost three spins is given in Eq.~(\ref{u3}) for the case of $a = 0$, and is that of applying an overall phase factor to all $a=0$ states.  

In summary, when applying any one of the operations shown in Fig.~\ref{u34} to the states (\ref{5spinsApp}) with $a=0$ or $b=0$, the result is either the identity operation or the multiplication by a phase factor that depends on only one of the qubit states.  These phases are simple to keep track of, and can be canceled by appropriate single-qubit rotations before or after the two-qubit gate sequence.  We can thus ignore the cases of $a=0$ or $b=0$, and work in the effective Hilbert space spanned by the states (\ref{effApp}).

\subsection{Matrix Representations in Pseudospin Space}
\label{matrixRepresentations}

To describe the action of the operations $U_3$, $\overline U_3$, and $U_4$ in the basis (\ref{effApp}), we introduce a corresponding pseudospin for both the $f=\frac12$ and $\frac32$ sectors.  Each pseudospin space is spanned by the states
\begin{eqnarray}
	\uparrow_f = (\blacktriangle(\bullet\blacktriangle)_{c=1/2})_{f}, \quad \downarrow_f = (\blacktriangle(\bullet\blacktriangle)_{c=3/2})_{f}.
	\label{pseudospinApp}
\end{eqnarray} 
For simplicity, below we refer to these spaces as $\uparrow_f$.  

The operation $U_4$ is shown in Fig.~\ref{u34}(a).  Its matrix representation is given most easily in its natural basis,
\begin{eqnarray}
	((\blacktriangle\blacktriangle)_d\bullet)_f,
	\label{dbasis}
\end{eqnarray}
which is related to the pseudospins $\uparrow_f$ further below.  If $f=\frac12$, we find from Eq.~(\ref{u4}) for the case of $b = 1$ with basis ordering $d=\{0, 1\}$,
\begin{eqnarray}
	U_{4, d}^{f=1/2}(\phi) = \text{diag}(1, e^{-i\phi}) = e^{-i \phi/2} e^{i \phi \axis z\cdot \boldsymbol\sigma/2}.
	\label{U4d1/2}
\end{eqnarray}
Here the subscript in the notation $U_{4, d}^{f=1/2}$ indicates the basis.  If $f=\frac32$, we similarly find with $d=\{1, 2\}$,
\begin{equation} 
	U_{4, d}^{f=3/2}(\phi)=\textnormal{diag}(e^{-i\phi},e^{-i\phi}) = e^{-i\phi}\mathbb{1}. % alternative: \mathds{1}.  
	\label{U4d3/2}
\end{equation}

We perform the basis change from the $d$-basis (\ref{dbasis}) to the pseudospin $\uparrow_f$ basis given in Eq.~(\ref{pseudospin}) in two steps.  First, we introduce another two pseudospin spaces spanned by the states
\begin{eqnarray}
	\uparrow'_f = (\blacktriangle(\blacktriangle\bullet)_{c=1/2})_{f}, \quad \downarrow'_f = (\blacktriangle(\blacktriangle\bullet)_{c=3/2})_{f}
	\label{pseudospin'}
\end{eqnarray} 
with $f=\frac12$ or $\frac32$,  which below we referred to as $\uparrow'_f$, and give the action of $U_4$ on this pseudospin.  Note that the $\uparrow'_f$ basis is related to the $d$-basis (\ref{dbasis}) by a basis change that consists of shifting ovals.  In the second step, we use the \textsc{powt} operation [see Eq.~(\ref{generalizedSWAP}) for $b=1$], which interchanges the particles $\blacktriangle = (\bullet\bullet)_1$ and $\bullet$, to relate the pseudospin bases $\uparrow'_f$ and $\uparrow_f$ to one another.  The action of \textsc{powt}, as discussed in Appendix \ref{PseudospinChange}, is to apply a relative phase factor of $-1$ between the states with $c=\frac12$ and $\frac32$.  This action can thus be interpreted as a $\pi$ rotation about the $z$ axis when relating $\uparrow'_f$ to $\uparrow_f$, i.e.,
\begin{eqnarray}
	U_{4, \uparrow_{f}} = e^{i \pi \axis z\cdot \boldsymbol\sigma/2}U_{4, \uparrow'_f}e^{i \pi \axis z\cdot \boldsymbol\sigma/2}
	\label{change_pseudpsins}
\end{eqnarray}  
for both the $f=\frac12$ and $\frac32$ sectors.  

Since the action of $U_4$ for $f=\frac32$ is proportional to the identity [cf.~Eq.~(\ref{U4d3/2})], here the two-step basis change has no effect,
\begin{eqnarray}
	U^{}_{4, \uparrow_{3/2}}(\phi) = U^{f=3/2}_{4, d}(\phi) = e^{-i\phi} \mathbb{1}.
	\label{f32}
\end{eqnarray} 
Changing bases in the $f=\frac12$ sector is less trivial.  The first step, in which we change from the $d$-basis (\ref{dbasis}) to the pseudospin $\uparrow'_{1/2}$ basis, has been carried out explicitly in Ref.~\cite{zeuch14} [see Eqs.~(17)-(21) therein] with the result that $U_4$ performs a pseudospin rotation
\begin{eqnarray}
	U_{4, \uparrow'_{1/2}}^{}(\phi) = e^{-i \phi/2} e^{i \phi \axis n^{\prime}_2\cdot \boldsymbol\sigma/2}
	\label{u4'}
\end{eqnarray}
about an axis $\axis n_2^{\prime}=(-2\sqrt{2}/3, 0, -1/3)$ through angle $\phi$.  The second step given in Eq.~(\ref{change_pseudpsins}), which is the transformation from $\uparrow'_f$ to $\uparrow_f$, has the effect of rotating the vector $\axis n_2^{\prime}$ about the $z$ axis through angle $\pi$ onto the vector $\axis n_2=(2\sqrt{2}/3,0,-1/3)$,
\begin{eqnarray}
	U_{4, \uparrow_{1/2}}^{}(\phi) &=& e^{i \pi \boldsymbol\sigma_z/2}U_{4, \uparrow'_{1/2}}^{}(\phi) e^{i \pi \boldsymbol\sigma_z/2}\nonumber \\
			&=& e^{-i \phi/2} e^{i \pi \boldsymbol\sigma_z/2}e^{i \phi \axis n^{\prime}_2\cdot \boldsymbol\sigma/2}e^{i \pi \boldsymbol\sigma_z/2} \nonumber \\
			&=& e^{-i \phi/2}e^{i \phi \axis n_2\cdot \boldsymbol\sigma/2}.
	\label{f12}
\end{eqnarray}
This $f=\frac12$ pseudospin rotation is indicated in Fig.~\ref{u34}(a).   

Next we determine the action of the two different $U_3$ operations shown in Figs.~\ref{u34}(b) and (c) on the pseudospin space $\uparrow_f$ given in Eq.~(\ref{pseudospinApp}).  To do this, we write both matrix representations in the basis (\ref{effApp}).  The operation $\overline U_3$ shown in Fig.~\ref{u34}(b) conserves the quantum number $c$, so we can take its matrix in the basis $c=\{\frac12, \frac32\}$ directly from Eq.~(\ref{u3bar}) for $a'=1$,
\begin{equation}
	\overline U_{3, c}(\phi) = \textnormal{diag}(1, e^{-i\phi}) = e^{-i\phi/2} e^{i\phi \axis z\cdot \boldsymbol\sigma /2},
	\label{u3barApp}
\end{equation}
which corresponds to a $z$-axis rotation in each pseudospin space $\uparrow_f$.  

The matrix representation of the $U_3$ operation in Fig.~\ref{u34}(c) can be found most easily in the basis $((\blacktriangle\bullet)_{c'}\blacktriangle)_{1/2}$.  For basis ordering $c'=\{\frac12,\frac32\}$, we find from Eq.~(\ref{u3}) for the case of $a=1$ that this matrix is the same as that given in Eq.~(\ref{u3barApp}).  Carrying out the basis change given below in Appendix \ref{explicit_calculations} [see Eqs.~(\ref{F3sum})-(\ref{F3}), in particular the definition of the matrix $F_3 = \axis f_3 \cdot \boldsymbol \sigma$ with $\axis f_3 = (2\sqrt{2}/\sqrt{3},0,-1/3)$], we find the matrix of $U_{3}$ in the basis (\ref{effApp}) with $c=\{\frac12,\frac32\}$,
\begin{equation}
	U_{3, c}(\phi) =  F_3\textnormal{diag}(1, e^{-i\phi})F_3 = e^{-i\phi/2} e^{i\phi \axis n_3\cdot \boldsymbol\sigma /2}
	\label{u3App}
\end{equation}
where $\axis n_3 = 2 \axis f_3 (\axis f_3 \cdot \axis z) - \axis z = (-4\sqrt 2/9, 0, -7/9)$.  The operation $\overline U_3$ is thus a rotation about $\axis n_3$ in each pseudospin space $\uparrow_f$.

% Appendix D
\section{Basis Changes and Qubit Rotations}
\label{explicit_calculations}

We now present further detailed derivations of matrix representations of unitary operations due to pulse sequences introduced above.  All required basis changes are presented in \ref{basis_changes}.  In \ref{four} we consider the $T$ sequence introduced in Sec.~\ref{22} for the case of the explicit pulse sequence given in Appendix \ref{TS}.  Finally, \ref{PseudospinChange} contains a derivation of the action of the \textsc{powt} operation introduced in Sec.~\ref{23}.

\subsection{Basis Changes}
\label{basis_changes}

We begin by describing a number of required basis changes.  First, consider three spin-$\frac12$ particles with total spin $\frac12$ in the standard qubit basis $(\bullet(\bullet\bullet)_a)_{1/2}$ with $a=0$ or 1, as also shown in Fig.~\ref{qubits}(a).  Another basis for this three-spin Hilbert space is $((\bullet\bullet)_{a'}\bullet)_{1/2}$ with $a'=0$ or 1, which can be related to the former basis via
\begin{eqnarray}
	((\bullet\bullet)_{a'}\bullet)_{1/2} = \sum_{a=0,1} F_{1, a'a}^{} (\bullet(\bullet\bullet)_a)_{1/2}
	\label{F1sum}
\end{eqnarray}
with the recoupling coefficients
\begin{eqnarray}
	F_{1, a'a}^{} =   \langle (\bullet(\bullet\bullet)_a)_{1/2} | ((\bullet\bullet)_{a'}\bullet)_{1/2}\rangle.
	\label{F1coefficients}
\end{eqnarray}
To fix the phases of these coefficients we must at this point adopt a phase convention for our total-spin basis states.  Here and in all that follows we use the standard Condon-Shortley convention \cite{messiah62}.  For this choice, the recoupling coefficients form the transformation matrix
\begin{eqnarray}
	F_1 = \left(\begin{array}{cc}
		-1/2 & \sqrt{3}/2 \\
		\sqrt{3}/2 & 1/2
	\end{array}\right)
	=
	\hat {\bf f}_1 \cdot \boldsymbol\sigma
	\label{F1}
\end{eqnarray}
with $\hat {\bf f}_1=(\sqrt{3}/2,0,-1/2)$ which maps the basis $a=\{0,1\}$ to the basis $a'=\{0,1\}$.  Note that $F_1^{\dagger} = F_1$.  

For the next basis change, consider four spin-$\frac12$ particles with total spin 1 in the basis $((\bullet\bullet)_{b'}(\bullet\bullet)_b)_1$, which is also shown in Fig.~\ref{Tfig}(b).  If we let $b\rightarrow 1$ and introduce an effective spin-$1$ particle $\blacktriangle = (\bullet\bullet)_{1}$ as in Eq.~(\ref{introduceBlackTriangle}), we obtain a two-dimensional, effective Hilbert space of two spin-$\frac12$ particles and one spin-1 particle with total spin 1.  We can now relate the basis $((\bullet\bullet)_{b'}\blacktriangle)_1$ with $b'=0$ and 1 to an alternate basis $(\bullet(\bullet\blacktriangle)_c)_1$ with $c=\frac12$ or $\frac32$,
\begin{eqnarray}
	(\bullet(\bullet\blacktriangle)_c)_{1} = \sum_{b'=0,1} F_{2, cb'}^{} ((\bullet\bullet)_{b'}\blacktriangle)_{1},
	\label{F2sum}
\end{eqnarray}
where the recoupling coefficients are given by
\begin{eqnarray}
	F_{2, cb'}^{} = \langle ((\bullet\bullet)_{b'}\blacktriangle)_{1/2} | (\bullet(\bullet\blacktriangle)_c)_{1/2} \rangle.
	\label{F2coefficients}
\end{eqnarray}
The transformation matrix 
\begin{eqnarray}
	F_2 = \left(\begin{array}{cc}
		-1/\sqrt3 & \sqrt{2/3} \\
		\sqrt{2/3} & 1/\sqrt3
	\end{array}\right)
	=
	\hat {\bf f}_2 \cdot \boldsymbol\sigma,
	\label{F2}
\end{eqnarray}
where $\hat {\bf f}_2=(\sqrt{2/3},0,-1/\sqrt{3})$ and $F_2^\dagger = F_2$,
then maps the basis $c=\{\frac12,\frac32\}$ to $b'=\{0,1\}$.  

Finally, we consider the basis change for the Hilbert space spanned by one spin-$\frac12$ and two spin-1 particles.  We begin from the basis $(\blacktriangle(\bullet\blacktriangle)_c)_{f=1/2}$ with $c=\frac12$ or $\frac32$, which is shown in Fig.~\ref{u34}.  Now consider the alternate basis $((\blacktriangle\bullet)_{c'}\blacktriangle)_{f=1/2}$ with $c'=\frac12$ and $\frac32$, which can be expressed in terms of the $c$-basis states by
\begin{eqnarray}
	((\blacktriangle\bullet)_{c'}\blacktriangle)_{1/2} = \sum_{c=\frac12, \frac32} F_{3, c'c}^{} (\blacktriangle(\bullet\blacktriangle)_c)_{1/2},
	\label{F3sum}
\end{eqnarray}
where
\begin{eqnarray}
	F_{3, c'c}^{} = \langle (\blacktriangle(\bullet\blacktriangle)_c)_{1/2} | ((\blacktriangle\bullet)_{c'}\blacktriangle)_{1/2} \rangle.
	\label{F3coefficients}
\end{eqnarray}
The transformation matrix
\begin{eqnarray}
	F_3 = \left(
	\begin{array}{cc}
		-1/3 & 2\sqrt 2/3 \\
		2\sqrt 2/3 & 1/3
	\end{array}
	\right)=
	\hat {\bf f}_3 \cdot \boldsymbol\sigma,
	\label{F3}
\end{eqnarray}
where $\hat {\bf f}_3 = (2\sqrt{2}/3,0,-1/3)$ and $F_3^{\dagger} = F_3$, then maps the basis $c=\{\frac12,\frac32\}$ to $c'=\{\frac12,\frac32\}$.

\subsection{$T$ Operation}
\label{four}

Figure \ref{observation1}(c) indicates that the $T$ operation is applied to a spin-$\frac12$ particle, $\bullet$, and an effective spin-$\frac12$ particle, $\bigstar$.  As is clear from Eq.~(\ref{unpacked}) and the ensuing discussion, the total-spin 1 Hilbert space that we have to consider (after unpacking the effective particle $\bigstar$) is three-dimensional, and it is spanned by the states
\begin{eqnarray}
	(\bullet(\bullet(\bullet\bullet)_b)_{c})_{d=1}
	\label{full_basis}
\end{eqnarray}
with $bc=0\frac12$, $bc=1\frac12$ and $bc=1\frac32$.  

An important condition placed on $T$ is that it conserve the quantum number $c$, which, being the total spin of the three-spin qubit with state label $b$ in Fig.~\ref{qubits}(b), is initialized to be $\frac12$.  We therefore seek the matrix representation of $T$ in what we here call the \emph{$T$-basis},
\begin{eqnarray}
	(\bullet(\bullet(\bullet\bullet)_b)_{c=1/2})_{d=1}, \qquad b = \{0, 1\},
	\label{desired_basis}
\end{eqnarray}
with the indicated basis ordering $b = \{0, 1\}$.  Labeling these four spins from left to right by 1 through 4 [see also Fig.~\ref{Tfig}(a)], the $T$ operation can be written as in Eq.~(\ref{T1}),
\begin{eqnarray}
	T = U_{34}(2s)V^{-1}U_{12}(t)U_{34}(t)V.
	\label{appendixT}
\end{eqnarray}
Here $s=2-t$, and the pulse sequence for $V$ is given in Fig.~\ref{Tfig}(d).  

We first consider the central two pulses within $T$, $U_{12}(t)U_{34}(t)$.  According to Fig.~\ref{Tfig}(b), the matrix representation of this operation (up to an overall phase) in the four-spin basis $((\bullet\bullet)_{b'}(\bullet\bullet)_b)_{1}$ with basis ordering $bb'=\{01,10,11\}$ is
\begin{equation}
	[U_{12}(t)U_{34}(t)]_{bb'} = e^{-i2\pi t}\text{diag}(e^{i\pi t},e^{i\pi t},1),
	\label{tt}
\end{equation}
Here the subscript in the notation $[U_{12}(t)U_{34}(t)]_{bb'}$ indicates evaluation in the $bb'$ basis.  

The $V$-sequence for the $T$-gate can, as indicated by the dashed line in Fig.~\ref{Tfig}(d), be formally divided into two parts,
\begin{eqnarray}
	V &=& U_{23}(\tfrac32)U_{12}(\tfrac12)U_{34}(\tfrac32)U_{23}(\tfrac12) \equiv V_0 V_1,
	\label{appendixV}
\end{eqnarray}
where $V_0 = U_{23}(\frac32)U_{12}(\frac12)$ and $V_1 = U_{34}(\frac32)U_{23}(\frac12)$.  Combining Eqs.~(\ref{appendixT}) and (\ref{appendixV}), we have
\begin{eqnarray}
	T = U_{34}(2s)V_1^{-1}V_0^{-1} U_{12}(t)U_{34}(t) V_0 V_1.
	\label{appendixT01}
\end{eqnarray}
To find the matrix representation of this $T$-operation (\ref{appendixT01}), we now perform a step-by-step evaluation of each of the operations surrounding the central two pulses $U_{12}(t)U_{34}(t)$.  

Given that the operation $V_0$ satisfies the constraint shown in Fig.~\ref{Tfig}(c), it maps a certain normalized superposition of the $c=\frac12$-states $(\bullet(\bullet(\bullet\bullet)_{b})_{c=1/2})_{1}$ with $b=0$ and 1 to the state
\begin{eqnarray}
	v_1 = ((\bullet\bullet)_{b'=1}(\bullet\bullet)_{b=1})_1.
	\label{v1}
\end{eqnarray}
Due to unitarity, the perpendicular superposition of the same $c=\frac12$-states is mapped into the $bb'=01,10$ sector onto the state
\begin{eqnarray}
	v_2 = \alpha ((\bullet\bullet)_1(\bullet\bullet)_0)_{1} + \beta ((\bullet\bullet)_0(\bullet\bullet)_1)_1
	\label{v2}
\end{eqnarray}
with $\alpha=-\frac{2+i}{\sqrt 6}$ and $\beta=\frac{i}{\sqrt{6}}$ (see also Note \cite{alphabeta}).  The action of the similarity transformation carried out by $V_0$ in Eq.~(\ref{appendixT}) can thus be viewed as a basis change from the $T$-basis (\ref{desired_basis}) to the $v$\emph{-basis} $\{v_1, v_2 \}$,
\begin{eqnarray}
	v_i = \sum_{b=0,1} F_0^{v_i b} (\bullet(\bullet(\bullet\bullet)_{b})_{1/2})_{1}.
	\label{activeBasisChange}
\end{eqnarray}
The coefficients $F_0^{v_ib}$ are given by
\begin{eqnarray}
	F_0^{v_ib} &=& \langle v_i| V_0 | (\bullet(\bullet(\bullet\bullet)_{b})_{1/2})_{1} \rangle.
	\label{F5coefficients}
\end{eqnarray}

To find the coefficients of $F_0$, we note
\begin{eqnarray}
	\langle v_1 | U_{23}(\tfrac32)U_{12}(\tfrac12) | (\bullet(\bullet(\bullet\bullet)_{b=0})_{1/2})_{1} \rangle = -\tfrac{1}{2}, \quad \label{F01}
	\\
	\langle v_1 | U_{23}(\tfrac32)U_{12}(\tfrac12) | (\bullet(\bullet(\bullet\bullet)_{b=1})_{1/2})_{1} \rangle = \tfrac{\sqrt{3}i}{2}. \quad 
	\label{F02}
\end{eqnarray}
These overlaps can be computed straightforwardly by evaluating $U_{12}$ in the basis $((\bullet\bullet)_{b'}(\bullet\bullet)_{b})_1$ and $U_{23}$ in the basis $(\bullet((\bullet\bullet)_{b''  }\bullet)_{c})_1$ using Eq.~(\ref{pulse}) (with an appropriate choice for the overall phase), together with Eqs.~(\ref{F1}) and (\ref{F2}).  Accordingly, the matrix
\begin{eqnarray}
	F_0 = \left(
	\begin{array}{cc}
		-1/2	& \sqrt{3}i/2\\
		-\sqrt{3}i/2 & 1/2
	\end{array}
	\right) = \hat {\bf f}_0\cdot\boldsymbol \sigma
	\label{F5}
\end{eqnarray}
with $\hat{\bf f}_0=(0,-\sqrt3/2,-1/2)$ maps the $T$-basis (\ref{desired_basis}) to the $v$-basis $\{v_1, v_2\}$.  [We note that this choice of $v$-basis ordering corresponds to $F_0=F_0^{\dagger}$.]

Notice that $U_{12}(t)U_{34}(t)$ as given in Eq.~(\ref{tt}) multiplies each state (\ref{v1}) and (\ref{v2}) by a certain phase factor; it is thus diagonal in the $v$-basis,
\begin{eqnarray}
	[U_{12}(t)U_{34}(t)]_{v_i} &=& e^{-i2\pi t}\text{diag}(1, e^{i\pi t}) \nonumber \\
			&=& e^{-i3\pi t/2}e^{-i\pi t \hat {\bf z}\cdot \boldsymbol\sigma/2}.
	\label{ttvi}
\end{eqnarray}
To examine the effect of the innermost similarity transformation due to $V_0$ in Eq.~(\ref{appendixT01}), we let
\begin{eqnarray}
	T_0=V_0^{-1} U_{12}(t)U_{34}(t) V_0.
	\label{appendixT0}
\end{eqnarray}
We evaluate this operation $T_0$ by carrying out the basis change (\ref{activeBasisChange}) to the $T$-basis,
\begin{eqnarray}
	T_{0, b}^{} &=& e^{-i3\pi t/2} \ F_0 e^{-i\pi t \hat {\bf z}\cdot \boldsymbol\sigma/2} F_0 \nonumber \\
			&=& e^{-i3\pi t/2}e^{-i\pi t\hat {\bf n}_0 \cdot \boldsymbol\sigma/2}, \qquad 
	\label{T0b-basis}
\end{eqnarray}
where 
\begin{eqnarray}
	\hat {\bf n}_0 = 2 (\hat {\bf f}_0\cdot \hat {\bf z}) \hat {\bf f}_0 - \hat {\bf z}=(0,\sqrt{3}/2,-1/2).
	\label{n0}
\end{eqnarray}

We note that the $T_0$-sequence (\ref{appendixT0}) for $t=1$ equals the $R$ sequence shown in Fig.~\ref{FWobservation1}(a), which has been used in deriving the Fong-Wandzura sequence \cite{zeuch16}.  According to the matrix representation (\ref{R}) of $R$ in the basis $(\bullet(\bullet(\bullet\bullet)_b)_{1/2})_{d} = (\bullet \bigstar)_{d}$, we have $R = M$ for $d=1$.  In this basis, which corresponds to the $T$-basis (\ref{desired_basis}), the matrix $M$ is then
\begin{eqnarray}
	M = T_{0, b}^{} {\big |}_{t=1} \stackrel{(\ref{T0b-basis})}{=} e^{-i3\pi/2}e^{-i\pi \hat {\bf n}_0 \cdot \boldsymbol\sigma/2} = \hat {\bf n}_0 \cdot \boldsymbol\sigma. \qquad
	\label{appendix_n0}
\end{eqnarray}

We now determine the matrix representations of each of the remaining pulses of the $T$-sequence (\ref{appendixT01}) in the $T$-basis (\ref{desired_basis}).  The operation $U_{34}(t)$ can be evaluated directly in this basis using Eq.~(\ref{pulsePseudo}),
\begin{eqnarray}
	U_{34, b}^{}(t) = e^{-i\pi t/2}e^{i\pi t\hat{\bf z}\cdot\boldsymbol\sigma/2}.
	\label{U34}
\end{eqnarray}
The action of $U_{23}(t)$ can be found directly in the basis $(\bullet((\bullet\bullet)_{b''}\bullet)_{1/2})_0$ with $b''=\{0,1\}$ using Eq.~(\ref{pulsePseudo}),
\begin{eqnarray}
	U_{23, b''}^{}(t) = e^{-i\pi t/2}e^{i\pi t\hat{\bf z}\cdot\boldsymbol\sigma/2}.
\end{eqnarray}
To find the matrix of this pulse in the $T$-basis, we carry out the basis change (\ref{F1sum}) to find
\begin{eqnarray}
	U_{23, b}^{}(t) = F_1 U_{23, b''}^{}(t) F_1.
	\label{U23}
\end{eqnarray}

Having determined every operation of the $T$-sequence in the $T$-basis (\ref{desired_basis}), we from now on work exclusively in this basis and drop the additional subscripts indicating the current basis.  To determine the effect of the outer similarity transformation in Eq.~(\ref{appendixT01}) due to $V_1=U_{34}(\frac32)U_{23}(\frac12)$, we first simplify $V_1$ as follows,
\begin{eqnarray}
	V_1 &=& U_{34}(\tfrac32)U_{23}(\tfrac12) \nonumber \\
		&\stackrel{(\ref{U34}),(\ref{U23})}=& e^{-i(\pi/2)\hat{\bf z}\cdot\boldsymbol\sigma/2} [F_1 e^{i(\pi/2)\hat{\bf z}\cdot\boldsymbol\sigma/2} F_1] \nonumber \\
		&=& [e^{-i(\pi/2)\hat{\bf z}\cdot\boldsymbol\sigma/2} F_1 e^{i(\pi/2)\hat{\bf z}\cdot\boldsymbol\sigma/2}] F_1 \nonumber \\
		&\equiv& F_4F_1,
\end{eqnarray}
so that
\begin{eqnarray}
	V_1^{-1} T_0 V_1 = F_1 F_4 T_0 F_4 F_1.
	\label{fun}
\end{eqnarray}
Here, the matrix $F_4 = e^{-i(\pi/2)\hat{\bf z}\cdot\boldsymbol\sigma/2} F_1 e^{i(\pi/2)\hat{\bf z}\cdot\boldsymbol\sigma/2}$
is the result of carrying out a similarity transformation on $F_1$,
\begin{eqnarray}
	F_4 = \hat{\bf f}_4\cdot \boldsymbol\sigma = \left(
	\begin{array}{cc}
		-1/2 & -i\sqrt{3}/2\\
		i\sqrt{3}/2 & 1/2
	\end{array}
	\right) = F_4^\dagger,
	\label{F6}
\end{eqnarray}
where $\hat{\bf f}_4=(0,\sqrt{3}/2,-1/2)$ is the result of rotating $\hat{\bf f}_1=(\sqrt{3}/2,0,-1/2)$ through an angle of $-\pi/2$ about the $z$ axis.  Note that $\hat {\bf f}_4 = \hat {\bf n}_0$, because of which the $F_4$ matrix in Eq.~(\ref{fun}) commutes with $T_0\sim e^{-i\pi t\hat {\bf n}_0 \cdot \boldsymbol\sigma/2}$ and thus has no effect, allowing us to simplify
\begin{eqnarray}
	V_1^{-1} T_0 V_1 &=& e^{-i3\pi t/2} [F_1 e^{-i\pi t\hat {\bf n}_0 \cdot \boldsymbol\sigma/2} F_1] \nonumber \\
			&=& e^{-i3\pi t/2}e^{-i\pi t\hat {\bf n}_1 \cdot \boldsymbol\sigma/2}
	\label{Talmost}
\end{eqnarray}
with $\hat {\bf n}_1 = 2(\hat {\bf n}_0 \cdot \hat {\bf f}_1)\hat {\bf f}_1 - \hat {\bf n}_0 = (\sqrt{3}/4,-\sqrt{3}/2,1/4)$.  The matrix representation of the operation due to the last pulse, $U_{34}(2s)$, is given in Eq.~(\ref{U34}), so that
\begin{eqnarray}
	T	&=& U_{34}(2s) V_1^{-1} T_0 V_1 \nonumber \\
		&=& [e^{-i2\pi s /2}e^{i2\pi s{\hat {\bf z}}\cdot \boldsymbol{\sigma}/2}] [e^{-i3\pi t/2}e^{-i\pi t\hat{\bf n}_1\cdot \boldsymbol \sigma/2}] \nonumber \\
		&=& e^{-i\pi t/2}e^{i\pi (2-t){\hat {\bf z}}\cdot \boldsymbol{\sigma}} e^{-i\pi t\hat{\bf n}_1\cdot \boldsymbol \sigma/2}
	\label{Tfinal}
\end{eqnarray}
where we used $s = 2-t$.  

Finally, associating this result (\ref{Tfinal}) given in the $T$-basis, that is $(\bullet(\bullet(\bullet\bullet)_b)_{1/2})_{d=1} = (\bullet \bigstar)_{d=1}$, with the matrix $\mathbb{M}$ in Eq.~(\ref{T}), we conclude
\begin{eqnarray}
	\mathbb{M} &=& e^{-i\pi t/2}e^{i\pi (2-t){\hat {\bf z}}\cdot \boldsymbol{\sigma}} e^{-i\pi t\hat{\bf n}_1\cdot \boldsymbol \sigma/2} \nonumber \\
			& \equiv &  e^{i\pi t/2} e^{i \phi(t) \hat {\bf n}(t) \cdot \boldsymbol \sigma/2} 
			\label{AppendixM}
\end{eqnarray}
with the effective rotation angle $\phi(t) = 2 \arccos((5\cos(\pi t/2)+3\cos(3\pi t/2))/8)$ and a unit vector $\hat {\bf n}(t)$.  Since $\phi(0) = 0$ and  $\phi(t_1 \equiv 4 \arctan(\sqrt{2-\sqrt{3}})) = 2\pi$ the pulse sequence constructed in Sec.~\ref{22} can thus be used to carry out arbitrary controlled-rotation gates using values of $t\in[0, t_1]$.

\subsection{Pseudospin Transformation}
\label{PseudospinChange}

In Appendix \ref{pseudospin_rotations} we use a transformation from one pseudospin space, which is spanned by the states
\begin{eqnarray}
	\uparrow_{1/2} = (\blacktriangle(\bullet\blacktriangle)_{c=1/2})_{1/2}, \quad \downarrow_{1/2} = (\blacktriangle(\bullet\blacktriangle)_{c=3/2})_{1/2},\quad \quad \quad
	\label{pseudospinApp2}
\end{eqnarray}
to another pseudospin space spanned by
\begin{eqnarray}
	\uparrow'_{1/2} = (\blacktriangle(\blacktriangle\bullet)_{c=1/2})_{1/2}, \quad \downarrow'_{1/2} = (\blacktriangle(\blacktriangle\bullet)_{c=3/2})_{1/2}.\quad \quad\quad
	\label{pseudospin'App2}
\end{eqnarray}
This change of bases is realized by interchanging the rightmost effective spin-1, $\blacktriangle$, and the spin-$\frac12$, $\bullet$.  In the present section we denote the corresponding operation, which is introduced in Sec.~\ref{23} as \textsc{powt}, by $U$.  

Taking into account that the total spin of these two particles, $c$, is conserved, and ignoring the leftmost $\blacktriangle$ in the above pseudospin states (which remains unchanged by this transformation), we write
\begin{eqnarray}
	U|(\bullet\blacktriangle)_{c}\rangle &=& F_{c} |(\blacktriangle\bullet)_{c}\rangle.
	\label{trafo}
\end{eqnarray}
The basis change from $(\bullet\blacktriangle)_{c}$ to $(\blacktriangle\bullet)_{c}$ with $c=\{\frac12, \frac32\}$ is thus characterized by a diagonal matrix $F = \text{diag}(F_{c=1/2}, F_{c=3/2})$.  

\begin{figure}
	\includegraphics[width=\columnwidth]{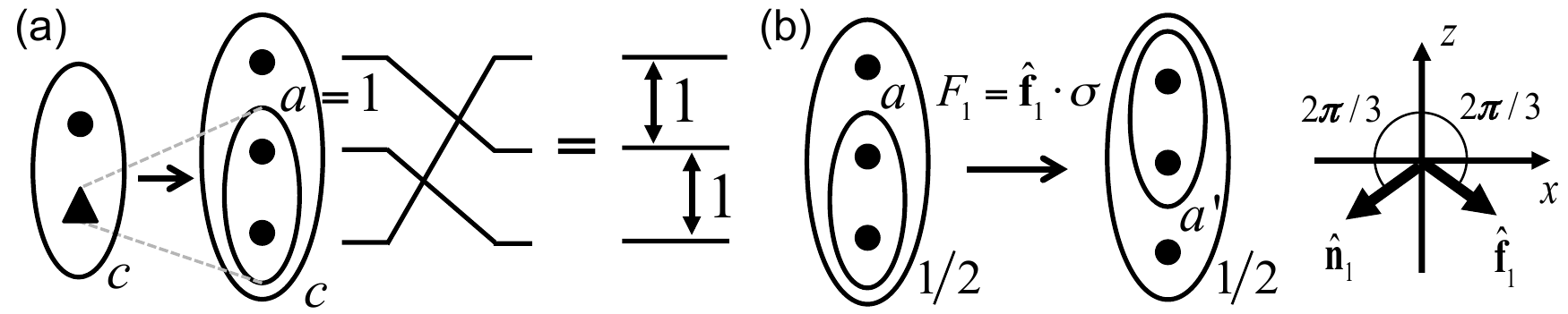}
	\caption{\textsc{powt} operation (\ref{trafo}) consisting of, (a), two regular \textsc{swap}s followed by, (b), basis change carried out by $F_1$.  In (a), the effective spin-1, $\blacktriangle$, is replaced with two spin-$\frac12$ particles, $\bullet$, with total spin $a=1$.}
	\label{swaps}
\end{figure}

Figure \ref{swaps} shows the basic information required to understand the basis change (\ref{trafo}).  As shown in Fig.~\ref{swaps}(a), we replace the spin-1 particle by two spin-$\frac12$ particles with total spin 1,
\begin{eqnarray}
	(\bullet \blacktriangle)_c &\rightarrow& (\bullet(\bullet\bullet)_{a=1})_c. \label{rep2}
\end{eqnarray}
The full Hilbert space of three spin-$\frac12$ particles is spanned by the states $(\bullet(\bullet\bullet)_a)_c$ with $ac=0\frac12$, $ac=1\frac12$ and $ac=1\frac32$.  Again referring to Fig.~\ref{swaps}(a), this basis change is then carried out by
\begin{eqnarray}
	U=U_{23}(1)U_{12}(1),
	\label{U}
\end{eqnarray}
where we labeled the spins 1 through 3 from left to right (or top to bottom in Fig.~\ref{swaps}).  The action of these \textsc{swap}s can be computed straightforwardly using simple spin wave functions; however, in the spirit of this work, we commit, as we have throughout, to using only total spin quantum numbers and recoupling coefficients, adopting the Condon-Shortley phase convention.   

The transformation (\ref{trafo}) now reads
\begin{eqnarray}
	U|(\bullet(\bullet\bullet)_{a=1})_c\rangle &=& F_{c} |((\bullet\bullet)_{a'=1}\bullet)_{c}\rangle,
	\label{trafo32}
\end{eqnarray}
where the coefficients are determined by
\begin{eqnarray}
	F_{c} &=& \langle ((\bullet\bullet)_{1}\bullet)_c | U | (\bullet(\bullet\bullet)_{1})_c\rangle.
	\label{Fcoefficients}
\end{eqnarray}
Note that the states on either side of this equation are given in different bases.  In the case of total spin $c=\frac32$ the transformation between these bases is trivial,
\begin{eqnarray}
	\langle ((\bullet\bullet)_{1}\bullet)_{3/2}| = \langle (\bullet(\bullet\bullet)_{1})_{3/2}|,
	\label{basisChange32}
\end{eqnarray}
so that here a \textsc{swap} acting on the two leftmost spins has the same effect as a \textsc{swap} acting on the two rightmost spins.  From Eq.~(\ref{pulse}) for the case of $a=1$ we find $U^{c=3/2}_{12}(1) = U^{c=3/2}_{23}(1) = e^{-i\pi} = -1$.  Using Eq.~(\ref{U}) we obtain for $c=\frac32$ that $U^{c=3/2} = (-1)^2 = 1$, so that
\begin{eqnarray}
	F_{c=3/2} = \langle (\bullet(\bullet\bullet)_{1})_{3/2} | U^{c=3/2} | (\bullet(\bullet\bullet)_{1})_{3/2} \rangle = 1. \quad
\end{eqnarray}

For the two-dimensional $c=\frac12$ sector using the matrix $F_1$ given in Eq.~(\ref{F1}), the $c=1/2$ equivalent of Eq.~(\ref{basisChange32}) is 
\begin{eqnarray}
	 \langle ((\bullet\bullet)_{1}\bullet)_{1/2}| = \sum_{a=0,1} F_{1,1a}\langle  (\bullet(\bullet\bullet)_{a})_{1/2}|,
	\label{basisChange12}
\end{eqnarray}
where we have used $F_1^\dagger = F_1$.  

Combining Eqs.~(\ref{Fcoefficients}) for $c=\frac12$ and (\ref{basisChange12}) we obtain
\begin{equation}
	F_{c=1/2} = \sum_{a=0,1} F_{1,1a}\langle  (\bullet(\bullet\bullet)_{a})_{1/2}| U^{c=1/2} | (\bullet(\bullet\bullet)_{1})_{1/2}\rangle.
	\label{Fcoefficients_c=1/2}
\end{equation}
% old version:  change basis of ket vector
%\begin{eqnarray}
%	F_{1/2} &=& \langle ((\bullet\bullet)_{1}\bullet)_{1/2} | U | (\bullet(\bullet\bullet)_{1})_{1/2}\rangle \nonumber \\
%		&=& \langle ((\bullet\bullet)_{1}\bullet)_{1/2} |  U_{23}(1)U_{12}(1)F_{1}| ((\bullet\bullet)_{1}\bullet)_{1/2} \rangle. \quad
%	\label{Fcoefficients_c=1/2}
%\end{eqnarray}
To find this matrix element, let us first determine the matrix representations of the two \textsc{swap}s making up $U$ for total spin $c=\frac12$.  The matrix of $U_{23}(1)$ in the basis $(\bullet(\bullet\bullet)_{a})_{1/2}$ with state ordering $a = \{0, 1\}$ is given by Eq.~(\ref{pulsePseudo}) for $t=1$, $U^{c=1/2}_{23}(1) = e^{-i\pi/2}e^{i\pi \axis z\cdot \boldsymbol\sigma/2}$.  
%\begin{eqnarray}
%	U^{c=1/2}_{23}(1) &=& e^{-i\pi/2}e^{i\pi \axis z\cdot \boldsymbol\sigma/2}.
%	\label{u12}
%\end{eqnarray}
Similarly, in the alternate basis $((\bullet\bullet)_{a'}\bullet)_{1/2}$ with $a' = \{0, 1\}$ the matrix of $U_{12}(1)$ is also given by Eq.~(\ref{pulsePseudo}) for $t=1$.  Changing from this $a'$ basis to the $a$ basis given above, we have $U^{c=1/2}_{12}(1) = F_1e^{-i\pi/2}e^{i\pi \axis z\cdot \boldsymbol\sigma/2}F_1$.  
%\begin{eqnarray}
%	U^{c=1/2}_{12}(1) = F_1e^{-i\pi/2}e^{i\pi \axis z\cdot \boldsymbol\sigma/2}F_1.
%	\label{u23}
%\end{eqnarray}

In the basis $a = \{0, 1\}$, the operator $F_1 U^{c=1/2}$ is then
\begin{eqnarray}
	F_1 U^{c=1/2} &=& F_1 e^{-i\pi/2}e^{i\pi \axis z\cdot \boldsymbol\sigma/2} (F_1e^{-i\pi/2}e^{i\pi \axis z\cdot \boldsymbol\sigma/2}F_1) \nonumber \\
%			&=& F_1 (e^{-i\pi/2}e^{i\pi \axis z\cdot \boldsymbol\sigma/2} F_1e^{-i\pi/2}e^{i\pi \axis z\cdot \boldsymbol\sigma/2})F_1 \nonumber \\
			&\equiv& F_1 F_1' F_1,
	\label{F1U}
\end{eqnarray}
where $F_1'= (e^{-i\pi/2}e^{i\pi \axis z\cdot \boldsymbol\sigma/2}) F_1 (e^{-i\pi/2}e^{i\pi \axis z\cdot \boldsymbol\sigma/2}) = \hat {\bf f}'_1 \cdot \boldsymbol\sigma$.  The vector $\hat {\bf f}_1'=(-\sqrt{3}/2,0,-1/2)$ is the result of rotating $\hat {\bf f}_1$ through $\pi$ about the $z$ axis.  With reference to Fig.~\ref{swaps}(b), we evaluate $F_1 F_1' F_1$ noting that rotating the vector $\hat {\bf f}_1'$ through $\pi$ about $\hat {\bf f}_1$ results in the vector $\hat {\bf z}$, implying that $F_1 U^{c=1/2} = e^{-i \pi/2} e^{i \pi \sigma_z/2} = \text{diag}(1, -1)$ in the usual basis ordering $a=\{0, 1\}$.  Equation (\ref{Fcoefficients_c=1/2}) thus yields
\begin{equation}
	F_{c=1/2} = \sum_{a=0,1} F_{1,1a}\langle  (\bullet(\bullet\bullet)_{a})_{1/2}| U^{c=1/2} | (\bullet(\bullet\bullet)_{1})_{1/2}\rangle = -1.
	\label{pseudospin_basis-change}
\end{equation}
We conclude that the basis change (\ref{trafo}) is carried out by the matrix
\begin{eqnarray}
  F &=& \text{diag}(F_{c=1/2}, F_{c=3/2}) = \text{diag}(-1, 1).
  \label{pseudospin-trafo}
\end{eqnarray}
This operation can be interpreted as a $z$-axis rotation through $\pi$ when performating the pseudpospin transformation from Eq.~(\ref{pseudospinApp2}) to (\ref{pseudospin'App2}) via the \textsc{powt} operation.

\bibliography{bibliography}

\begin{thebibliography}{48}
\bibitem [{opt()}]{optimal}%
  \BibitemOpen
  \href@noop {} {}\bibinfo {note} {The definition of an optimal pulse sequence
  is ambiguous. Here, by optimal we mean requiring the least number of pulses
  for spins oriented along a linear array, assuming the exchange coupling can
  only be turned on between nearest neighbors.}\BibitemShut {Stop}%
\bibitem [{n0()}]{n0}%
  \BibitemOpen
  \href@noop {} {}\bibinfo {note} {In Appendix \ref{four}, the vector $\axis
  n_0$ is explicitly calculated for the Condon-Shortley phase convention, see
  Eqs.~(\ref{n0}) and (\ref{appendix_n0}).}\BibitemShut {Stop}%
\bibitem [{gau()}]{gauge_freedom}%
  \BibitemOpen
  \href@noop {} {}\bibinfo {note} {The vector $\axis n_2$, and all vectors in
  the various pseudospin spaces discussed in this paper, can be rotated about
  the $z$ axis by multiplying the total-spin basis states by different phase
  factors. To fix this gauge degree of freedom we adopt the standard
  Condon-Shortley phase convention, as discussed in Appendix
  \ref{basis_changes}.}\BibitemShut {Stop}%
\bibitem [{V0-()}]{V0-is-enough}%
  \BibitemOpen
  \href@noop {} {}\bibinfo {note} {We note that the constraint for the $T$
  operation posed by Eq.~(\ref{T}), because of which we use the non-optimal $V$
  sequence shown in Fig.~\ref{Tfig}(d), can be relaxed by letting $T=(\mathbb
  M_1, \mathbb M_2)$, given in the same basis as Eq.~(\ref{T}). If $\mathbb M_1
  \mathbb M_2^{-1} \neq e^{i\delta}\mathbb{1}$ for some phase $\delta$, the
  corresponding sequence shown in Fig.~\ref{observation1}(c) enacts a
  leakage-free, entangling two-qubit gate. The shortest sequence for $T$ of the
  form (\ref{T1}) is then $T = V_0^{-1}U_{12}(t)U_{34}(t)V_0$. However, one can
  show that either choice of $T$ yields the same two-qubit gate sequence up to
  single-qubit rotations.}\BibitemShut {Stop}%
\bibitem [{alp()}]{alphabeta}%
  \BibitemOpen
  \href@noop {} {}\bibinfo {note} {The values for $\alpha$ and $\beta$, given
  here for completeness, are not relevant for our discussion, since the
  operation $U_{12}(t)U_{34}(t)$ acts as the identity in the $bb'd=011$, 101
  sector, see Eq.~(\ref{tt}).}\BibitemShut {Stop}%
\end{thebibliography}

\end{document}